\documentclass[pdflatex,sn-mathphys-num]{sn-jnl}

\usepackage{graphicx}
\usepackage{amsmath,amssymb,amsfonts}
\usepackage{amsthm}
\usepackage{booktabs}
\usepackage{siunitx}
\usepackage{subcaption}
\usepackage{algorithm}
\usepackage{algpseudocode}
\usepackage{placeins}

\makeatletter
\@ifundefined{orcidlogo}
  {\newcommand{\orcidlogo}{}}
  {\renewcommand{\orcidlogo}{}}
\makeatother

\begin{document}

\title[QFT]
{Interaction with the Environment via Random Matrices and the Emergence of Classical Fields}

\author*[1]{\fnm{Alexey A.} \sur{Kryukov}}\email{kryukov@uwm.edu}

\affil*[1]{\orgdiv{Department of Mathematics \& Natural Sciences},
\orgname{University of Wisconsin--Milwaukee},
\orgaddress{\city{Milwaukee}, \state{WI}, \country{USA}}}

\abstract{It was recently shown that Newtonian dynamics of macroscopic particles can be derived from unitary Schr\"odinger evolution under a random-matrix assumption on the system-environment interaction. In that framework, classical phase space is realized geometrically as a manifold of localized equivalence classes in quantum state space, the tangent component of Schr\"odinger evolution reproduces Newtonian motion, and environmental interactions stabilize the state near this manifold.
We extend this framework to quantum fields. The field itself is not assumed to become classical. Instead, macroscopic particles stabilized near the classical particle manifold interact with the field through the sector of field state space accessible to localized particle dynamics. The classical field is represented by the corresponding localized sector, and finite probe resolution leads to a quotient description in terms of localized equivalence classes of field states. The tangent component of the quantum-field Schr\"odinger dynamics on this localized quotient sector yields the corresponding classical field equations.
Finite-dimensional simulations illustrate the mechanism for scalar and electromagnetic fields. The accessible field coordinates satisfy the sourced Klein--Gordon and Maxwell equations, and a localized test charge responds to the electromagnetic field through the Lorentz force. Thus classical field behavior emerges within unitary Schr\"odinger dynamics, without identifying the classical field with an expectation value, without relying on coherent states as special physical states, and without introducing a nonunitary collapse postulate.}


\maketitle

\section{Introduction}

A classical electromagnetic field is observed through its interaction with macroscopic charges. Thus the question of how a quantum field appears classical should not be formulated as the question of whether the field state itself collapses onto a classical field configuration. Rather, the question is why the interaction between a quantum field and a macroscopic charge can be described by a classical field acting on that charge. In the framework developed here, the field remains quantum. What appears classical is the quotient sector of the field state space accessible to macroscopic particles whose own states are localized by an appropriate dynamical mechanism.

In previous work \cite{KryukovPHLA,KryukovPhysicsA,KryukovArx,KryukovCompArx}, Newtonian dynamics of macroscopic particles was derived from Schr\"odinger evolution by restricting the dynamics to a manifold of localized states in projective Hilbert space. The tangent component of the Schr\"odinger flow on this manifold reproduces the classical equations of motion. The same work introduced conjecture {\bf (RM)}, according to which environmental interactions of macroscopic systems may be modeled, during short interaction intervals, by random Hamiltonians drawn from the Gaussian Unitary Ensemble. This random-matrix dynamics repeatedly returns the state to a narrow neighborhood of the localized particle manifold and thereby stabilizes classical particle motion.

The present paper extends this idea to quantum fields. A macroscopic particle stabilized near the localized particle manifold does not probe arbitrary directions in the field Hilbert space. Because its own state is localized, its interaction with the field is sensitive, to leading order, only to the field directions coupled to the particle's localized motion. These are precisely the directions that define the tangential field dynamics observed by the particle. The finite spatial width of the localized state then adds a further coarse-graining: the particle couples to smeared field observables over its localization region. This leads naturally to a quotient description. Field functionals that agree, within the probe resolution, on the smeared observables accessible to localized macroscopic particles are grouped into the same equivalence class. The corresponding quotient-level field data are what enter the classical field description.

The classical field is therefore not an expectation value assigned to an arbitrary field state, nor is it the result of selecting coherent states as special physical states. In the present formulation, a classical field is represented by the localized sector of field-state space seen by macroscopic probes, and its dynamics is the tangent dynamics induced on that sector by the quantum-field Schr\"odinger evolution. Finite localization width then gives an additional quotient description: field states that agree, within the probe resolution, on the accessible smeared observables belong to the same localized equivalence class. Gaussian field functionals may be used as convenient representatives for computing the induced Fubini--Study geometry, but the physically relevant object is the localized sector, or equivalently its quotient into localized equivalence classes, not the particular representative.

We begin with finite-dimensional simulations. The scalar-field simulation shows how a localized probe selects field data from a quantum field and how the corresponding quotient-level coordinate satisfies the sourced Klein--Gordon equation. We then perform the analogous electromagnetic simulation, showing that the selected field coordinates satisfy Maxwell's equations, and use a localized test charge to verify that these same coordinates act through the Lorentz force. After the simulations, we give the geometric construction of the localized field sector, derive the associated tangent dynamics, and explain the role of Gaussian representatives and the limitations of coherent-state or Ehrenfest arguments. Throughout the paper we assume the \({\bf (RM)}\) stabilization mechanism for macroscopic sources and probes, as developed in \cite{KryukovPHLA,KryukovPhysicsA,KryukovArx,KryukovCompArx}; we also give a new estimate showing that the cosmic microwave background can provide an ultimate environmental source of such \({\bf (RM)}\)-type interactions for macroscopic bodies even in otherwise empty space.


\section{A scalar-field simulation}

We begin with a simple simulation in the functional Schr\"odinger representation. The purpose is to illustrate how a macroscopic localized particle probes a quantum field and thereby observes an effective classical field. The electromagnetic field will be treated separately.

There is an important operational point already in this scalar model. A macroscopic body does not measure a point value of a field. In their analysis of the measurability of electromagnetic field quantities, Bohr and Rosenfeld showed that the quantities accessible to macroscopic test bodies are field components averaged over finite spacetime regions \cite{BohrRosenfeld1933,BohrRosenfeld1979}. Their problem was the measurability of electromagnetic field quantities within quantum electrodynamics. The problem considered here is different: we ask how an underlying quantum field gives rise to an effective classical field when probed by macroscopic particles whose states are localized by the random-matrix environmental dynamics of \cite{KryukovPHLA,KryukovPhysicsA,KryukovArx,KryukovCompArx}. Nevertheless, the Bohr--Rosenfeld analysis supports the operational assumption used below: the field accessed by a macroscopic probe is a smeared field observable, not a pointwise field value.

The inner products used below should be understood in this geometric sense. They are not expectation values assigned to an arbitrary quantum field state and then declared classical. Rather, they compute coordinates labeling the localized equivalence class selected by the interaction with a macroscopic probe. A localized macroscopic particle couples only to the field observables determined by its localized profile, and hence only to the corresponding localized quotient sector of the field state space. The simulations below illustrate the resulting quotient-coordinate dynamics in simple finite-dimensional models. The full geometric explanation, in which this dynamics is identified with the tangential component of the quantum-field Schr\"odinger flow, is given in Section~\ref{sec:tangent}.




Let \(\mathbb P(\mathcal H_N)\) denote the projective Hilbert space of the
finite-dimensional field approximation, with field coordinates
\[
(\phi_1,\ldots,\phi_N)
\]
and conjugate momentum operators
\[
\widehat\pi_j
=
-i\hbar\frac{\partial}{\partial \phi_j}.
\]
For a normalized representative \(\Psi\) of a ray
\([\Psi]\in\mathbb P(\mathcal H_N)\), define
\[
\Phi_j(\Psi)
=
\langle \Psi,\widehat\phi_j\Psi\rangle,
\qquad
\Pi_j(\Psi)
=
\langle \Psi,\widehat\pi_j\Psi\rangle.
\]

Fix a localization width \(\Sigma_\phi\) for the field-configuration
variables. Let
\(\mathcal L_{\Sigma_\phi}\subset\mathbb P(\mathcal H_N)\) be the set of
projective states localized in field-configuration space, in the sense that
\[
\operatorname{Cov}_{\Psi}(\widehat\phi)\le \Sigma_\phi^2 I
\]
as positive-semidefinite matrices. Equivalently,
\[
v^T\operatorname{Cov}_{\Psi}(\widehat\phi)v
\le
\Sigma_\phi^2 |v|^2
\qquad
\text{for all }v\in\mathbb R^N .
\]
Here
\[
\widehat\phi
=
(\widehat\phi_1,\ldots,\widehat\phi_N),
\]
and \(\operatorname{Cov}_{\Psi}(\widehat\phi)\) denotes the covariance
matrix of the field-coordinate operators in the state \(\Psi\):
\[
\left(\operatorname{Cov}_{\Psi}(\widehat\phi)\right)_{jk}
=
\langle \Psi,\widehat\phi_j\widehat\phi_k\Psi\rangle
-
\Phi_j(\Psi)\Phi_k(\Psi),
\qquad
j,k=1,\ldots,N .
\]

Two localized projective states
\[
[\Psi],[\chi]\in \mathcal L_{\Sigma_\phi}
\]
are equivalent,
\[
[\Psi]\sim[\chi],
\]
if and only if
\[
\Phi_j(\Psi)=\Phi_j(\chi),
\qquad
\Pi_j(\Psi)=\Pi_j(\chi),
\qquad
j=1,\ldots,N.
\]
Thus the equivalence class labeled by
\[
(\Phi,\Pi)
=
(\Phi_1,\ldots,\Phi_N,\Pi_1,\ldots,\Pi_N)
\]
is
\[
[(\Phi,\Pi)]_{\Sigma_\phi}
=
\left\{
[\Psi]\in\mathcal L_{\Sigma_\phi}:
\Phi_j(\Psi)=\Phi_j,\ 
\Pi_j(\Psi)=\Pi_j,\ 
j=1,\ldots,N
\right\}.
\]
The quotient manifold of localized field states is defined by
\[
\mathcal M^{\Sigma_\phi}_{\rm field}
=
\mathcal L_{\Sigma_\phi}/\sim
=
\left\{
[(\Phi,\Pi)]_{\Sigma_\phi}:
(\Phi,\Pi)\in\mathbb R^{2N}
\right\}.
\]
The numbers
\[
(\Phi,\Pi)
=
(\Phi_1,\ldots,\Phi_N,\Pi_1,\ldots,\Pi_N)
\]
are coordinates on \(\mathcal M^{\Sigma_\phi}_{\rm field}\). They
label the equivalence class itself and are not tied to any particular
representative of the class.

For the simulation, we consider a real scalar field in one spatial dimension.
After discretization on a lattice with sites \(x_1,\ldots,x_N\), a field configuration is described by the vector
\[
\phi=(\phi_1,\ldots,\phi_N),
\]
where \(\phi_j\) is the value of the field at \(x_j\). The field state is a wave function
\[
\Psi(\phi_1,\ldots,\phi_N,t),
\]
which is the finite-dimensional version of a wave functional \(\Psi[\phi,t]\). The field Hamiltonian is taken to be
\[
\widehat H_\Phi
=
-\frac{\hbar^2}{2}
\sum_{j=1}^N
\frac{\partial^2}{\partial \phi_j^2}
+
\frac{\omega^2}{2}
\sum_{j=1}^N
\phi_j^2
+
\frac{\kappa}{2}
\sum_{j=1}^{N-1}
(\phi_{j+1}-\phi_j)^2 .
\]
The last term is the lattice analogue of the field-gradient energy. Thus this is a finite-dimensional approximation to the usual functional Schr\"odinger Hamiltonian for a scalar field.

Let the macroscopic particle be stabilized near the lattice point \(x_a\). Since the particle is macroscopic, its state is assumed to be already localized near the classical manifold by the {\bf (RM)} dynamics of \cite{KryukovPHLA,KryukovPhysicsA,KryukovArx,KryukovCompArx}. Therefore the primary simulation does not begin with a superposition of particle positions. Rather, we fix the particle location \(a\) and study the evolution of the quantum field seen by a probe localized near \(x_a\).

Because the particle has a finite spatial extent and finite resolution, its interaction with the field is described by a smeared field operator. In the finite-dimensional approximation this operator is
\[
\phi_W(a)
=
\sum_{j=1}^N W_j(a)\phi_j ,
\]
where the weights \(W_j(a)\) are concentrated around the particle center \(a\) and satisfy
\[
\sum_{j=1}^N W_j(a)=1.
\]
The corresponding field coordinate on the localized quotient manifold is
\[
\Phi_W^{(a)}(t)
=
\left\langle
\Psi_a(t),
\phi_W(a)\Psi_a(t)
\right\rangle
=
\sum_{j=1}^N W_j(a)\Phi_j^{(a)}(t).
\]
In the continuum limit the smeared field operator becomes
\[
\phi_W(a)
=
\int W_a(x)\phi(x)\,dx .
\]
The point value \(\phi(a)\) is only an idealization obtained in the limit in which the spatial support of \(W_a\) becomes arbitrarily small.

For fixed particle location \(a\), the corresponding field state \(\Psi_a\) evolves by
\[
i\hbar \frac{\partial \Psi_a}{\partial t}
=
\widehat H_a\Psi_a,
\]
where
\[
\widehat H_a
=
\widehat H_\Phi+\lambda \phi_W(a).
\]
Thus the particle location and spatial resolution determine which smeared field operator enters the interaction. We will see that the field does not split in advance into classical and non-classical components. Rather, a macroscopic probe accesses the field through a smeared field operator, and hence through the corresponding field coordinate, determined by the probe's localized state.

For the two-site model, \(N=2\), we take
\[
\phi_W(1)=W_1(1)\phi_1+W_2(1)\phi_2,
\qquad
\phi_W(2)=W_1(2)\phi_1+W_2(2)\phi_2 .
\]
In the numerical run below we use
\[
\phi_W(1)=0.85\phi_1+0.15\phi_2,
\qquad
\phi_W(2)=0.15\phi_1+0.85\phi_2 .
\]
The special choice \(W_j(a)=\delta_{ja}\) gives the idealized point-site interaction, but the smeared interaction is the physically relevant one.

For the present quadratic finite-dimensional model, the dynamical evolution of the relevant conditional field coordinates can be computed directly from the Schr\"odinger equation, or, in this quadratic model, by formulas equivalent to the Ehrenfest identities. For \(N=2\), the field Hamiltonian is
\[
\widehat H_\Phi
=
-\frac{\hbar^2}{2}
\left(
\frac{\partial^2}{\partial \phi_1^2}
+
\frac{\partial^2}{\partial \phi_2^2}
\right)
+
\frac{\omega^2}{2}(\phi_1^2+\phi_2^2)
+
\frac{\kappa}{2}(\phi_2-\phi_1)^2 .
\]
The induced equations for the field coordinates \(\Phi_j(t)\) and their conjugate momentum coordinates \(\Pi_j(t)\) are
\[
\dot \Phi_j=\Pi_j,
\]
and
\[
\dot\Pi_1
=
-\omega^2\Phi_1-\kappa(\Phi_1-\Phi_2)-\lambda W_1(a),
\]
\[
\dot\Pi_2
=
-\omega^2\Phi_2-\kappa(\Phi_2-\Phi_1)-\lambda W_2(a).
\]
Equivalently,
\[
\ddot\Phi_1+\omega^2\Phi_1+\kappa(\Phi_1-\Phi_2)
=
-\lambda W_1(a),
\]
\[
\ddot\Phi_2+\omega^2\Phi_2+\kappa(\Phi_2-\Phi_1)
=
-\lambda W_2(a).
\]
These are the two-site sourced Klein--Gordon equations. In the continuum limit, the discrete coupling term becomes the spatial Laplacian term, and the equation takes the form
\[
\partial_t^2 \Phi(x,t)-\partial_x^2\Phi(x,t)+m^2\Phi(x,t)
=
-\lambda W_a(x).
\]
Thus the field visible to the macroscopic probe satisfies the classical sourced Klein--Gordon equation, with the probe profile \(W_a\) acting as the source.


\subsection{Numerical implementation}

We now describe the numerical realization of the two-site scalar-field model. The field state is represented in the variables \((\phi_1,\phi_2)\). The initial field state is chosen to be a normalized Gaussian representative
\[
\Psi_0(\phi_1,\phi_2)
=
C
\exp\left[
-\frac{(\phi_1-\bar\phi_1)^2}{4s^2}
-\frac{(\phi_2-\bar\phi_2)^2}{4s^2}
+
\frac{i}{\hbar}
(\bar\pi_1\phi_1+\bar\pi_2\phi_2)
\right],
\]
where \(\bar\phi_j\) and \(\bar\pi_j\) are the initial field-position and conjugate-momentum coordinates. The parameter \(s\) is the localization width of the representative in field-configuration space and is assumed to be small compared with the scale on which the relevant field coordinates vary. Thus the state belongs to the localized equivalence class with field-position coordinates
\[
\Phi=(\bar\phi_1,\bar\phi_2)
\]
and conjugate-momentum coordinates
\[
\Pi=(\bar\pi_1,\bar\pi_2).
\]

For each fixed particle center \(a\), the field evolves under the corresponding Hamiltonian
\[
\widehat H_a
=
\widehat H_\Phi+\lambda \phi_W(a).
\]
Thus
\[
\Psi_a(t+\Delta t)
=
\exp\left[
-\frac{i\Delta t}{\hbar}
\widehat H_a
\right]\Psi_a(t).
\]
The evolution is computed by a split-operator method. Since \(\widehat H_a\) is quadratic in the field variables plus a linear source term, the induced equations for the quotient coordinates close in this finite-dimensional model.

At each time step we record the quotient coordinates
\[
\Phi_j^{(a)}(t)
=
\langle \Psi_a(t),\phi_j\Psi_a(t)\rangle,
\]
and
\[
\Pi_j^{(a)}(t)
=
\left\langle
\Psi_a(t),
-i\hbar\frac{\partial}{\partial \phi_j}\Psi_a(t)
\right\rangle .
\]
For each fixed probe profile \(W_a\), the interaction Hamiltonian selects the field coordinate
\[
\Phi_W^{(a)}(t)
=
\left\langle
\Psi_a(t),
\phi_W(a)\Psi_a(t)
\right\rangle .
\]
Equivalently,
\[
\Phi_W^{(a)}(t)
=
\sum_{j=1}^N W_j(a)\Phi_j^{(a)}(t).
\]

One may also consider an initial superposition of particle centers,
\[
\alpha |a_1\rangle+\beta |a_2\rangle,
\]
when describing a measurement of the particle while it interacts with the field. In that case the joint state has the form
\[
\Psi(t)
=
\alpha |a_1\rangle\Psi_{a_1}(t)
+
\beta |a_2\rangle\Psi_{a_2}(t),
\]
where the two conditional field states evolve under different Hamiltonians. This is not the primary situation considered in the present simulation. Here the macroscopic test particle is assumed to be already stabilized near the classical manifold by the {\bf (RM)} dynamics of \cite{KryukovPHLA,KryukovPhysicsA,KryukovArx,KryukovCompArx}. The measurement-related case, in which random-matrix stabilization acts on a superposition of particle centers, will be considered separately below.

\subsection{Simulation results}

We implemented the two-site scalar-field simulation for two fixed probe locations, \(a_1\) and \(a_2\). These are two separate macroscopic-probe runs, not a single run with a macroscopic particle in a superposition. The corresponding Hamiltonians are
\[
\widehat H_{a_1}
=
\widehat H_\Phi+\lambda \phi_W(a_1),
\qquad
\widehat H_{a_2}
=
\widehat H_\Phi+\lambda \phi_W(a_2).
\]
The numerical parameters are listed in Table~\ref{tab:scalar_sim_parameters}.

\begin{table}[h]
\centering
\begin{tabular}{c|c|l}
\hline
Parameter & Value & Description \\
\hline
\(\hbar\) & \(1\) & Planck constant in simulation units \\
\(\omega\) & \(1\) & local oscillator frequency of the field coordinates \\
\(\kappa\) & \(0.35\) & coupling between neighboring field sites \\
\(\lambda\) & \(0.75\) & particle-field coupling strength \\
\(W(a_1)\) & \((0.85,0.15)\) & field-coordinate weights selected by a particle centered at \(a_1\) \\
\(W(a_2)\) & \((0.15,0.85)\) & field-coordinate weights selected by a particle centered at \(a_2\) \\
\hline
\end{tabular}
\caption{Parameters used in the two-site scalar-field simulation.}
\label{tab:scalar_sim_parameters}
\end{table}

We used an asymmetric initial field packet with coordinates
\[
\bar\phi_1(0)=0.45,
\qquad
\bar\phi_2(0)=-0.20,
\qquad
\bar\pi_1(0)=\bar\pi_2(0)=0.
\]
The initial selected field coordinates for the probe profiles \(W_{a_1}\) and \(W_{a_2}\) are
\[
\Phi_W^{(a_1)}(0)
=
0.3525,
\qquad
\Phi_W^{(a_2)}(0)
=
-0.1025.
\]
Thus two different probe profiles select different field coordinates already at the initial time. The initial and final values of these coordinates are shown in Table~\ref{tab:asymmetric_smeared_values}. The numerical values depend on the chosen parameters and initial state; their role is to show that different probe profiles correspond to different field coordinates selected by the interaction Hamiltonian.

\begin{table}[h]
\centering
\begin{tabular}{c|c|c}
\hline
Field coordinate & Initial value & Final value \\
\hline
\(\Phi_W^{(a_1)}\) & \(0.3525\) & \(-0.183845\) \\
\(\Phi_W^{(a_2)}\) & \(-0.1025\) & \(-0.141844\) \\
\hline
\end{tabular}
\caption{Initial and final values of the selected field coordinates corresponding to the probe profiles \(W_{a_1}\) and \(W_{a_2}\).}
\label{tab:asymmetric_smeared_values}
\end{table}

\begin{figure}[h]
\centering
\includegraphics[width=0.75\textwidth]{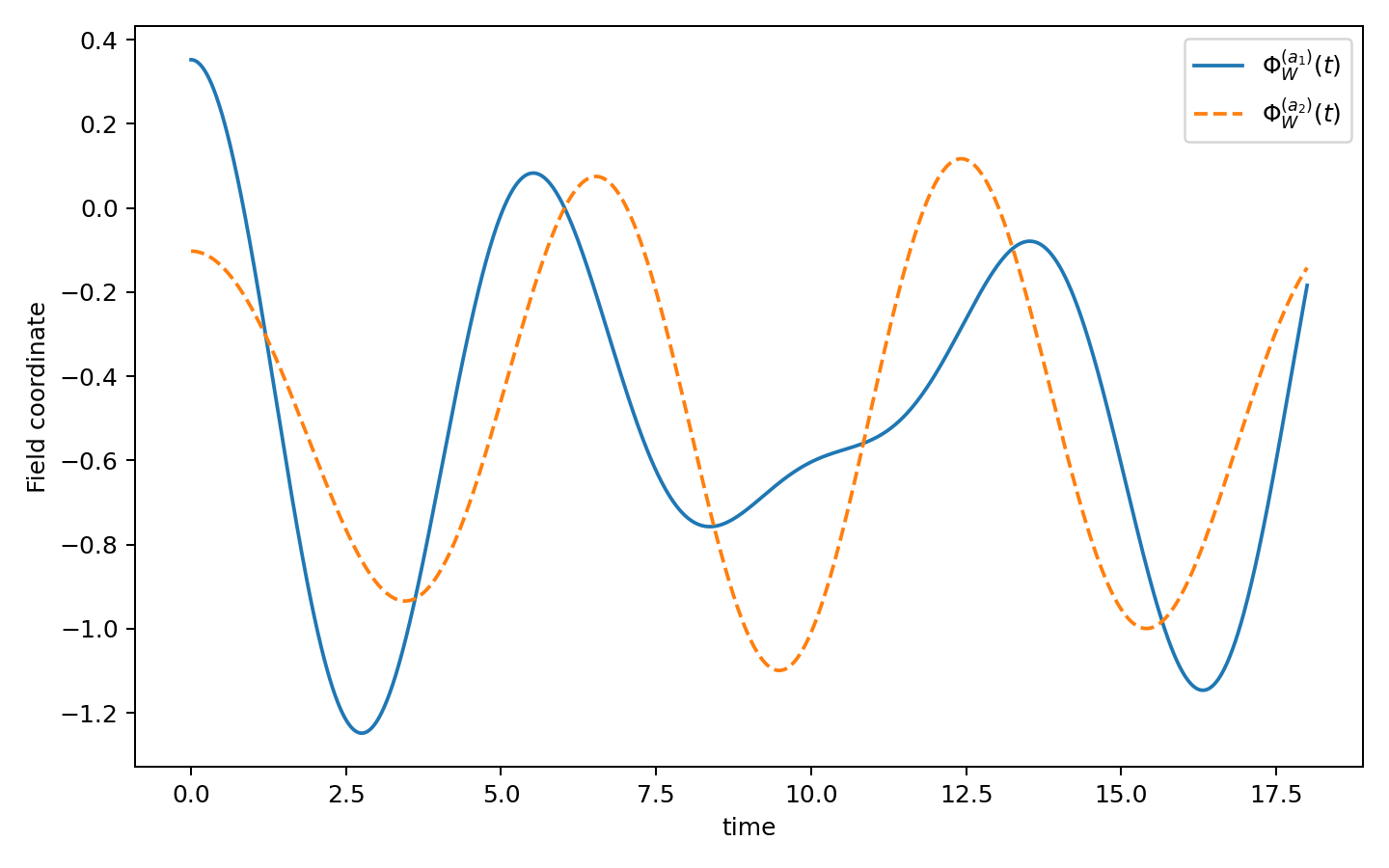}
\caption{Selected field coordinates \(\Phi_W^{(a_1)}(t)\) and \(\Phi_W^{(a_2)}(t)\) corresponding to the probe profiles \(W_{a_1}\) and \(W_{a_2}\).}
\label{fig:asymmetric_smeared_field}
\end{figure}

Figure~\ref{fig:asymmetric_smeared_field} shows the selected field coordinates
\(
\Phi_W^{(a_1)}(t)
=
\sum_j W_j(a_1)\Phi_j^{(a_1)}(t)
\)
and
\(
\Phi_W^{(a_2)}(t)
=
\sum_j W_j(a_2)\Phi_j^{(a_2)}(t).
\)

\begin{figure}[h]
\centering
\includegraphics[width=0.75\textwidth]{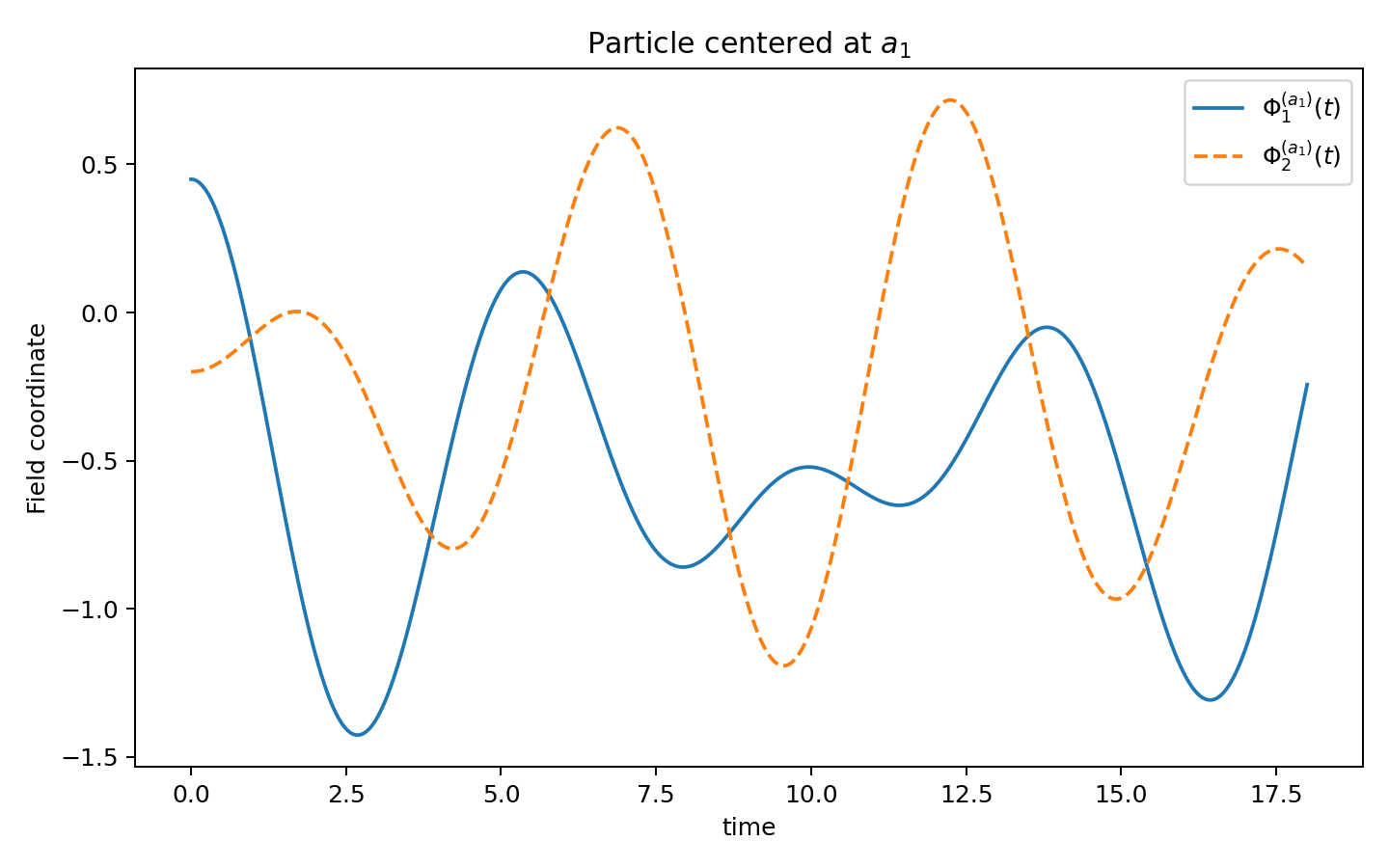}
\caption{Field-position coordinates \(\Phi_1^{(a_1)}(t)\) and \(\Phi_2^{(a_1)}(t)\) for the field evolution corresponding to the probe profile \(W_{a_1}\).}
\label{fig:coordinates_probe_x1}
\end{figure}

\begin{figure}[h]
\centering
\includegraphics[width=0.75\textwidth]{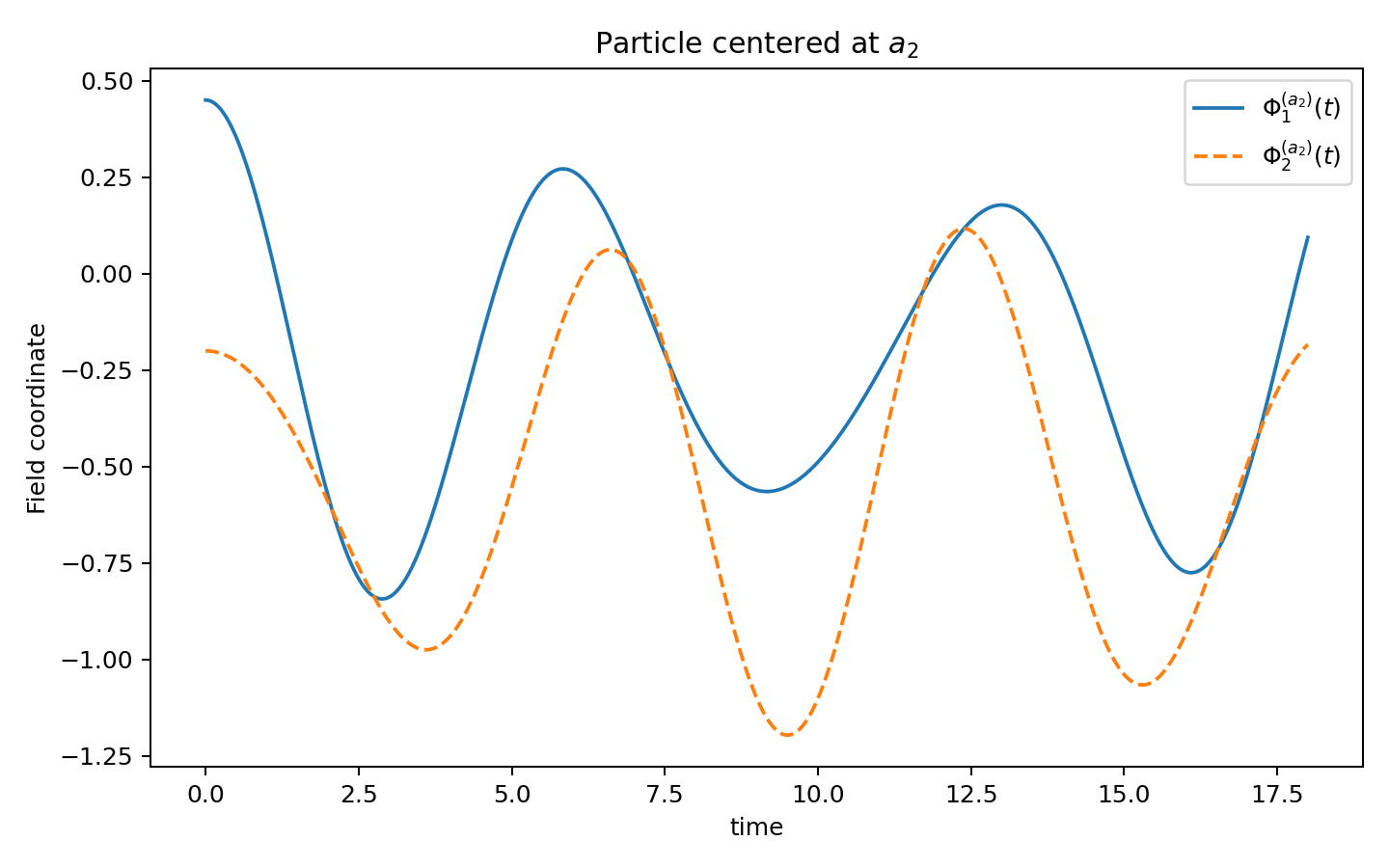}
\caption{Field-position coordinates \(\Phi_1^{(a_2)}(t)\) and \(\Phi_2^{(a_2)}(t)\) for the field evolution corresponding to the probe profile \(W_{a_2}\).}
\label{fig:coordinates_probe_x2}
\end{figure}

Figures~\ref{fig:coordinates_probe_x1} and~\ref{fig:coordinates_probe_x2} show the corresponding field-position coordinates \(\Phi_1^{(a)}(t)\) and \(\Phi_2^{(a)}(t)\) for 
\(a=a_1\) and \(a_2\).

To verify explicitly that the selected field coordinate obeys the classical equation, we compared the quotient-coordinate evolution obtained from the Schr\"odinger equation with the solutions of the two-site sourced Klein--Gordon equations
\[
\ddot\Phi_1+\omega^2\Phi_1+\kappa(\Phi_1-\Phi_2)
=
-\lambda W_1(a),
\]
\[
\ddot\Phi_2+\omega^2\Phi_2+\kappa(\Phi_2-\Phi_1)
=
-\lambda W_2(a),
\]
using the same initial data. The comparison is performed separately for the two fixed probe profiles \(W_{a_1}\) and \(W_{a_2}\), corresponding to \(a=a_1\) and \(a=a_2\).

For the probe profile \(W_{a_1}\), the maximum difference between the selected field coordinate obtained from the Schr\"odinger evolution and the corresponding classical solution was
\[
1.90\times 10^{-4},
\]
with RMS difference
\[
5.69\times 10^{-5}.
\]
For the probe profile \(W_{a_2}\), the maximum difference was
\[
8.69\times 10^{-5},
\]
with RMS difference
\[
4.30\times 10^{-5}.
\]
The remaining discrepancy is due to numerical time-stepping and grid error.

\begin{table}[h]
\centering
\begin{tabular}{c|c|c}
\hline
Particle center & Maximum difference & RMS difference \\
\hline
\(a_1\) & \(1.90\times 10^{-4}\) & \(5.69\times 10^{-5}\) \\
\(a_2\) & \(8.69\times 10^{-5}\) & \(4.30\times 10^{-5}\) \\
\hline
\end{tabular}
\caption{Differences between the field-coordinate values obtained from the Schr\"odinger evolution and the corresponding classical solutions of the two-site sourced Klein--Gordon equations.}
\label{tab:kg_comparison_errors}
\end{table}

\begin{figure}[h]
\centering
\includegraphics[width=0.75\textwidth]{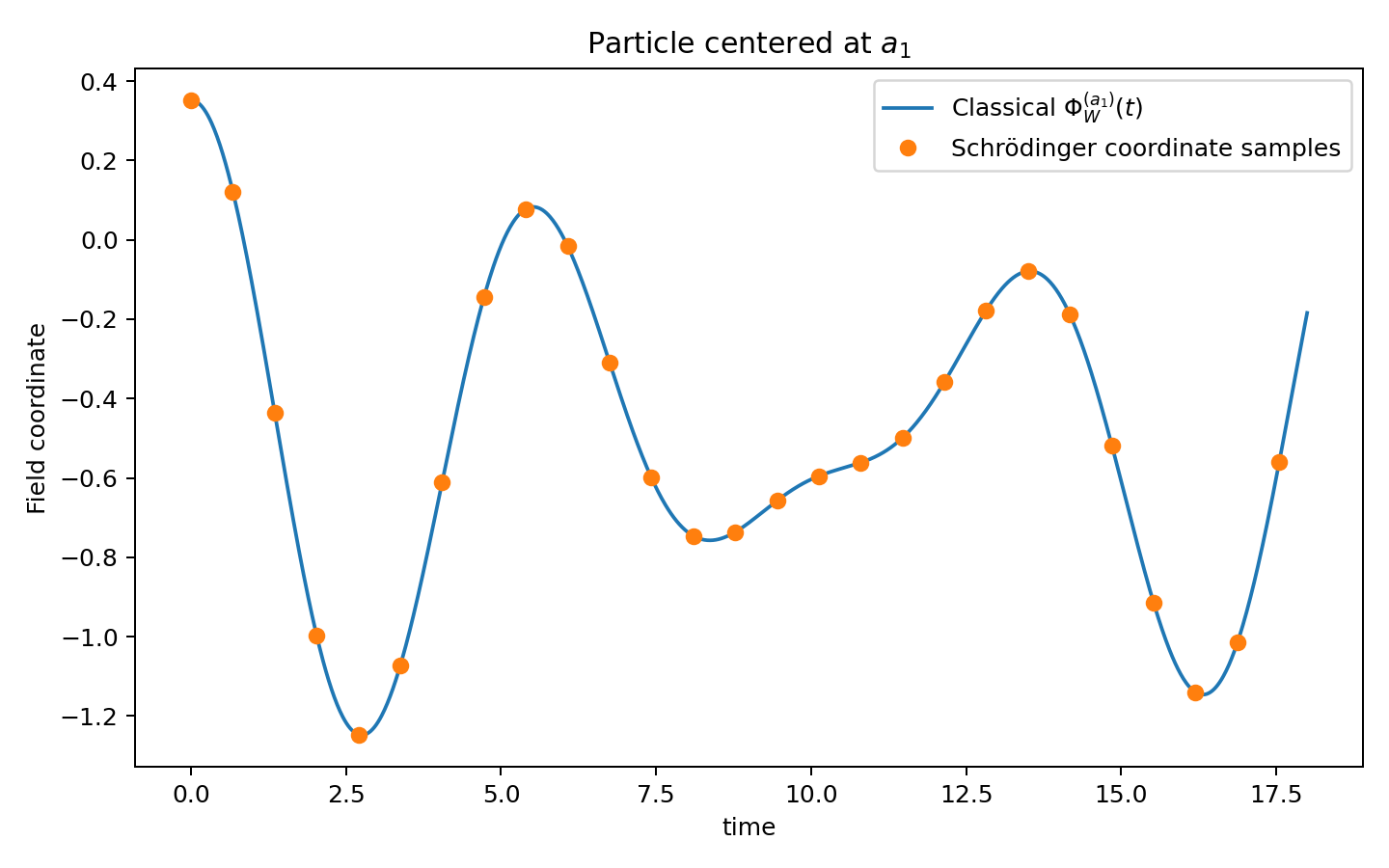}
\caption{The continuous curve is the classical solution for the field coordinate \(\Phi_W^{(a_1)}(t)\). The dots are sampled values of the same coordinate computed from the Schr\"odinger evolution. The agreement is within numerical error.}
\label{fig:kg_comparison_probe_x1}
\end{figure}

\begin{figure}[h]
\centering
\includegraphics[width=0.75\textwidth]{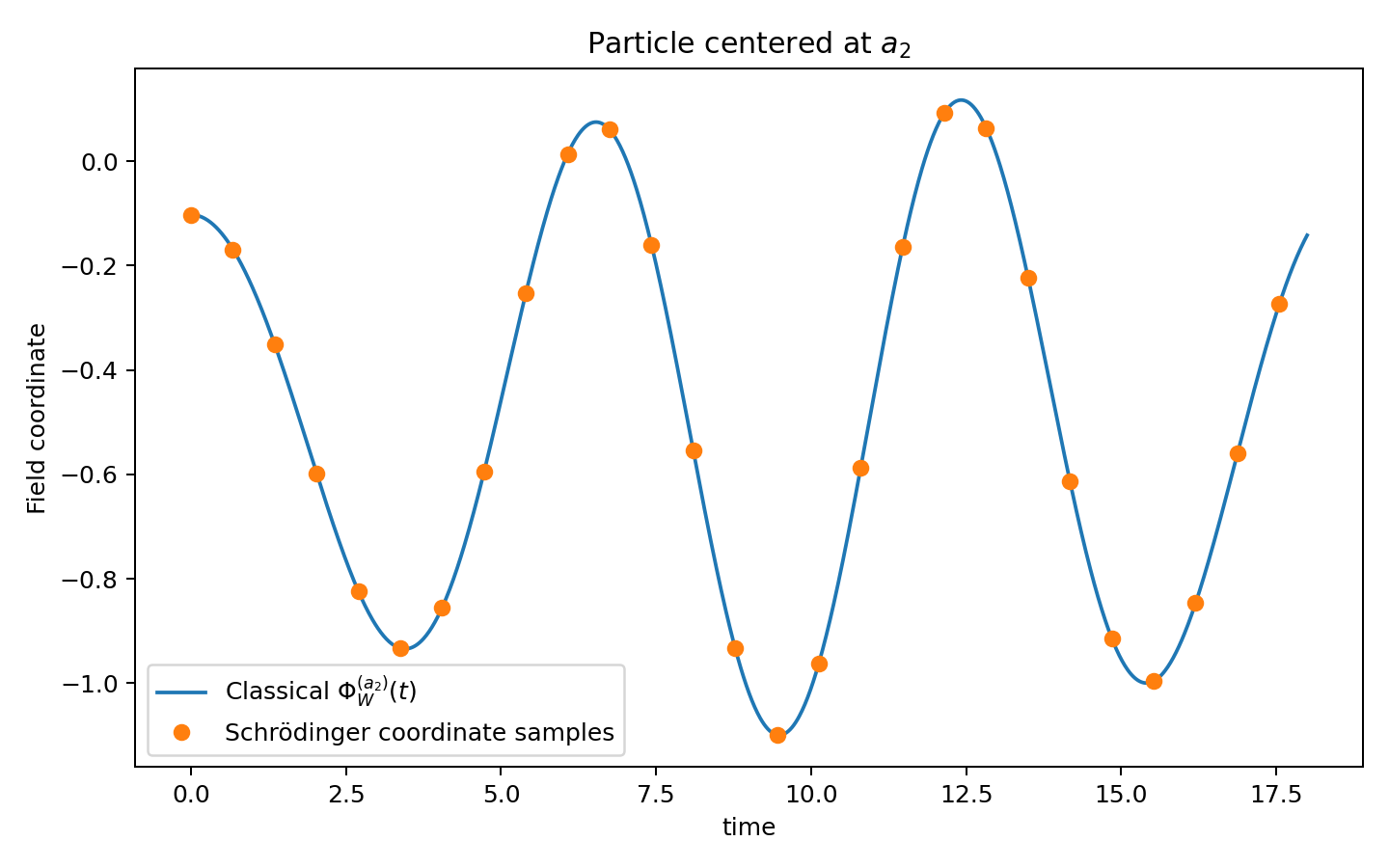}
\caption{The continuous curve is the classical solution for the field coordinate \(\Phi_W^{(a_2)}(t)\). The dots are sampled values of the same coordinate computed from the Schr\"odinger evolution. The agreement is within numerical error.}
\label{fig:kg_comparison_probe_x2}
\end{figure}

In Figures~\ref{fig:kg_comparison_probe_x1} and~\ref{fig:kg_comparison_probe_x2}, the continuous curve represents the classical solution of the two-site sourced Klein--Gordon equations. The dots are sampled values of the same coordinate obtained from the Schr\"odinger evolution. The dots lie on the classical curve up to numerical error.

These numerical results support the operational interpretation used in the paper. A localized macroscopic particle does not probe the full quantum field state and does not observe a point value of the field. Instead, for a particle with probe profile \(W_a\), the interaction Hamiltonian selects the field coordinate
\[
\Phi_W^{(a)}(t)
=
\left\langle
\Psi_a(t),
\phi_W(a)\Psi_a(t)
\right\rangle
=
\sum_j W_j(a)\Phi_j^{(a)}(t).
\]
This coordinate obeys the discrete sourced Klein--Gordon equation, and in the continuum limit it satisfies
\[
\partial_t^2 \Phi(x,t)-\partial_x^2\Phi(x,t)+m^2\Phi(x,t)
=
-\lambda W_a(x).
\]
The simulation therefore illustrates, in a finite-dimensional setting, how the quotient-coordinate dynamics of a quantum field reproduces the classical sourced field equation relative to a localized macroscopic probe.

\section{An electromagnetic-field simulation}

We now repeat the same finite-dimensional mechanism for the electromagnetic field. The purpose is not yet to describe the motion of a test charge in a given field, but rather to show that the field coordinates accessible to a macroscopic current source satisfy the classical Maxwell equations. The Lorentz-force response of a test charge will be considered separately.

To avoid unnecessary gauge complications, we begin with a one-dimensional transverse electromagnetic field. Let \(A(x,t)\) be a transverse component of the vector potential. In the functional Schr\"odinger representation, a quantum state of the field is a wave functional
\[
\Psi[A,t].
\]
The corresponding classical Hamiltonian, in suitable units, is
\[
H_{\mathrm{EM}}
=
\frac12
\int
\left[
\Pi(x)^2
+
c^2(\partial_x A(x))^2
\right]dx ,
\]
where \(\Pi(x)\) is the momentum conjugate to \(A(x)\). In the quantum theory, \(\Pi(x)\) is represented by the functional derivative operator
\[
\widehat\Pi(x)
=
-i\hbar\frac{\delta}{\delta A(x)}.
\]

A macroscopic current localized near \(a\) does not couple to a point value of the vector potential. Rather, as in the scalar-field case, it couples to a smeared field coordinate
\[
A_W(a)
=
\int W_a(x)A(x)\,dx ,
\]
where \(W_a\) is concentrated near the position of the macroscopic source. For a prescribed current amplitude \(J_a(t)\), the interaction Hamiltonian is
\[
H_{\mathrm{int}}
=
-
J_a(t)A_W(a).
\]
For a fixed probe profile \(W_a\), the interaction with the prescribed current \(J_a(t)\) gives the sourced field Hamiltonian
\[
H_a
=
H_{\mathrm{EM}}
-
J_a(t)A_W(a).
\]
After discretization on a lattice \(x_1,\ldots,x_N\), the field configuration is
\[
A=(A_1,\ldots,A_N),
\]
and the smeared vector potential is
\[
A_W(a)
=
\sum_{j=1}^N W_j(a)A_j .
\]
The discretized Hamiltonian is
\[
\widehat H_a
=
\frac12
\sum_j
\left[
-\hbar^2\frac{\partial^2}{\partial A_j^2}
+
c^2
\left(
\frac{A_{j+1}-A_j}{\Delta x}
\right)^2
\right]\Delta x
-
J_a(t)
\sum_j W_j(a)A_j\Delta x .
\]
The conditional field wave function \(\Psi_a(A_1,\ldots,A_N,t)\) therefore evolves by
\[
i\hbar\frac{\partial\Psi_a}{\partial t}
=
\widehat H_a\Psi_a .
\]

Let
\[
\mathcal A_j(t)
=
\langle \Psi_a(t),A_j\Psi_a(t)\rangle
\]
denote the vector-potential coordinate on the quotient field manifold selected by the sourced Hamiltonian. Since the Hamiltonian is quadratic in the field variables and linear in the source, the Schr\"odinger equation gives a closed system for these coordinates. Equivalently, one may obtain it from the Ehrenfest equations. The result is the discrete sourced wave equation
\[
\ddot{\mathcal A}_j
=
c^2\Delta_d \mathcal A_j
+
J_a(t)W_j(a),
\]
where \(\Delta_d\) is the discrete Laplacian. In the continuum limit this becomes
\[
\partial_t^2 \mathcal A(x,t)
-
c^2\partial_x^2\mathcal A(x,t)
=
J_a(t)W_a(x).
\]
This is the one-dimensional transverse Maxwell wave equation with a smeared current source.

With the conventions
\[
\mathcal E(x,t)=-\partial_t\mathcal A(x,t),
\qquad
\mathcal B(x,t)=\partial_x\mathcal A(x,t),
\]
the wave equation is equivalent to the one-dimensional transverse Maxwell system
\[
\partial_t\mathcal B
=
-\partial_x\mathcal E,
\]
and
\[
\partial_t\mathcal E
=
-c^2\partial_x\mathcal B
-
J_a(t)W_a(x).
\]
Thus the conditional field averages visible to the macroscopic source obey the classical Maxwell equations.

\subsection{Numerical implementation}

We simulated the discretized transverse electromagnetic field with \(N=180\)
lattice points on an interval of length \(L=30\). The boundary condition was
\(A=0\) at the endpoints. The macroscopic current source was chosen to be a
localized oscillatory pulse of the form
\[
J(x,t)
=
J_0 W_a(x)
\sin(\Omega t)
\exp\left[
-\frac{(t-t_0)^2}{2\tau^2}
\right],
\]
where \(W_a(x)\) is a normalized Gaussian profile with \(a=0\). The current
pulse generates outgoing transverse electromagnetic waves. The parameters are
listed in Table~\ref{tab:em_sim_parameters}.

\begin{table}[h]
\centering
\begin{tabular}{c|c|l}
\hline
Parameter & Value & Description \\
\hline
\(c\) & \(1\) & speed of light in simulation units \\
\(L\) & \(30\) & length of the spatial interval \\
\(N\) & \(180\) & number of lattice points \\
\(J_0\) & \(1\) & current amplitude \\
\(a\) & \(0\) & center of the macroscopic current source \\
\(\sigma_x\) & \(0.65\) & spatial width of the smearing function \(W_a\) \\
\(t_0\) & \(3.2\) & center time of the current pulse \\
\(\tau\) & \(0.9\) & temporal width of the current pulse \\
\(\Omega\) & \(3.2\) & oscillation frequency of the current pulse \\
\hline
\end{tabular}
\caption{Parameters used in the transverse electromagnetic-field simulation.}
\label{tab:em_sim_parameters}
\end{table}

The interaction of this current with the electromagnetic field is through the
smeared vector-potential operator
\[
\widehat A_J(t)
=
\int J(x,t)\widehat A(x)\,dx .
\]
Thus the current does not couple to arbitrary directions in the field Hilbert
space. It selects the field coordinate associated with the current profile
\(J(x,t)\). Field states that agree, within the resolution of the source, on
the corresponding smeared observable belong to the same source-selected
equivalence class.

For the numerical simulation, we choose a convenient representative of this
source-selected localized class. We take it to be Gaussian:
\[
\Psi_0[A]
=
C
\exp\left[
-\frac12
\int A(x)K(x,y)A(y)\,dx\,dy
\right].
\]
Here \(K(x,y)\) is a positive symmetric kernel, interpreted as the inverse
covariance kernel of the Gaussian representative. It determines the localization
width and correlations of the representative in the field variables selected by
the source profile. In the finite-dimensional simulation, \(K\) is replaced by a
positive symmetric matrix acting on the lattice field variables. The Gaussian
form is used only as a convenient representative of the localized equivalence
class selected by the macroscopic current; it is not an assumption that the
full electromagnetic field state collapses to a Gaussian functional.

The corresponding initial quotient coordinates are
\[
\mathcal A(x,0)=0,
\qquad
\Pi_{\mathcal A}(x,0)=0 .
\]

\begin{figure}[h]
\centering
\includegraphics[width=0.75\textwidth]{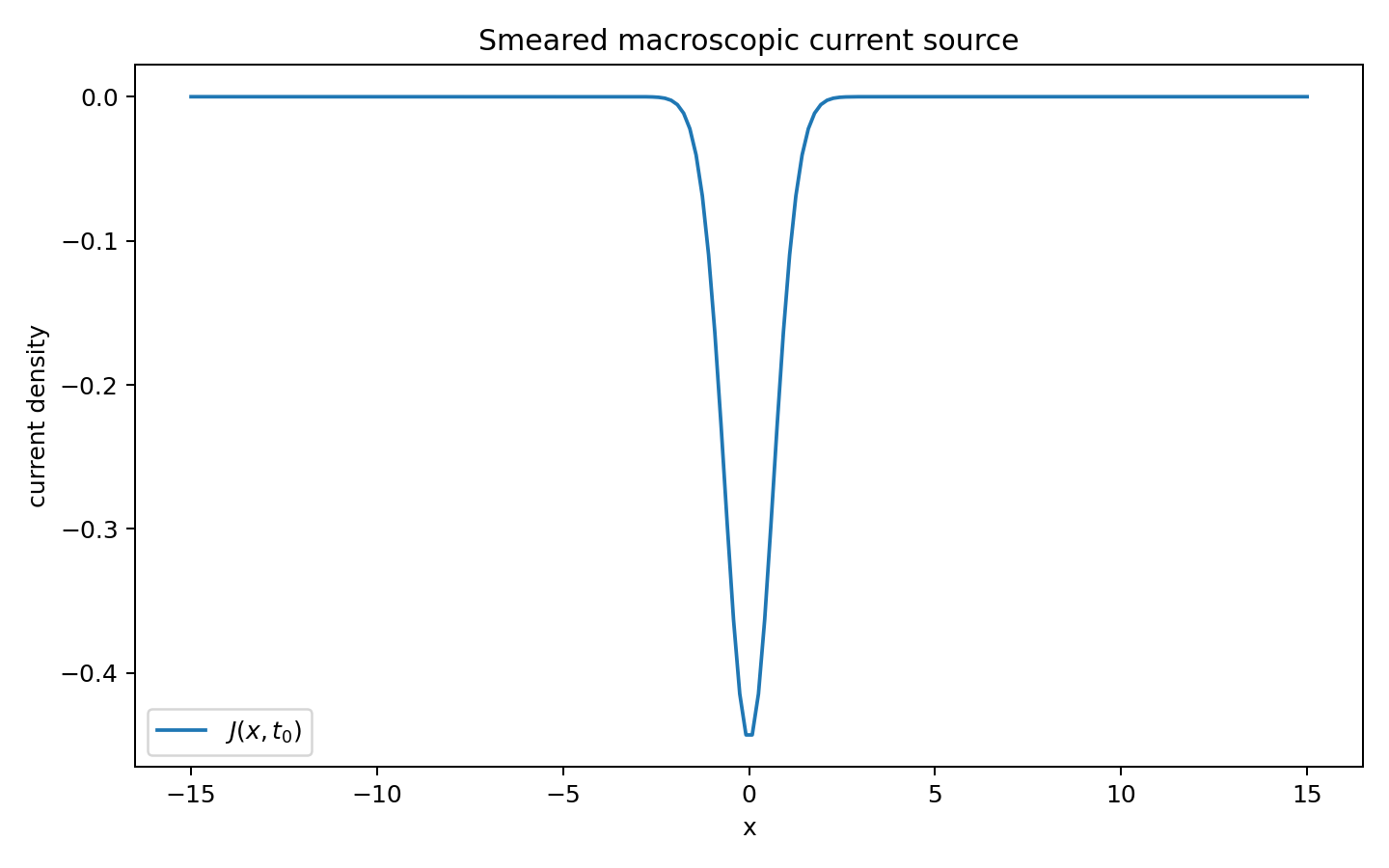}
\caption{Smeared current source \(J(x,t_0)\). The source is localized near \(a=0\) and represents the finite spatial extent of the macroscopic current profile.}
\label{fig:em_current_source}
\end{figure}

Figure~\ref{fig:em_current_source} shows the smeared current source \(J(x,t_0)\) at the peak time of the pulse.

\begin{figure}[h]
\centering
\includegraphics[width=0.75\textwidth]{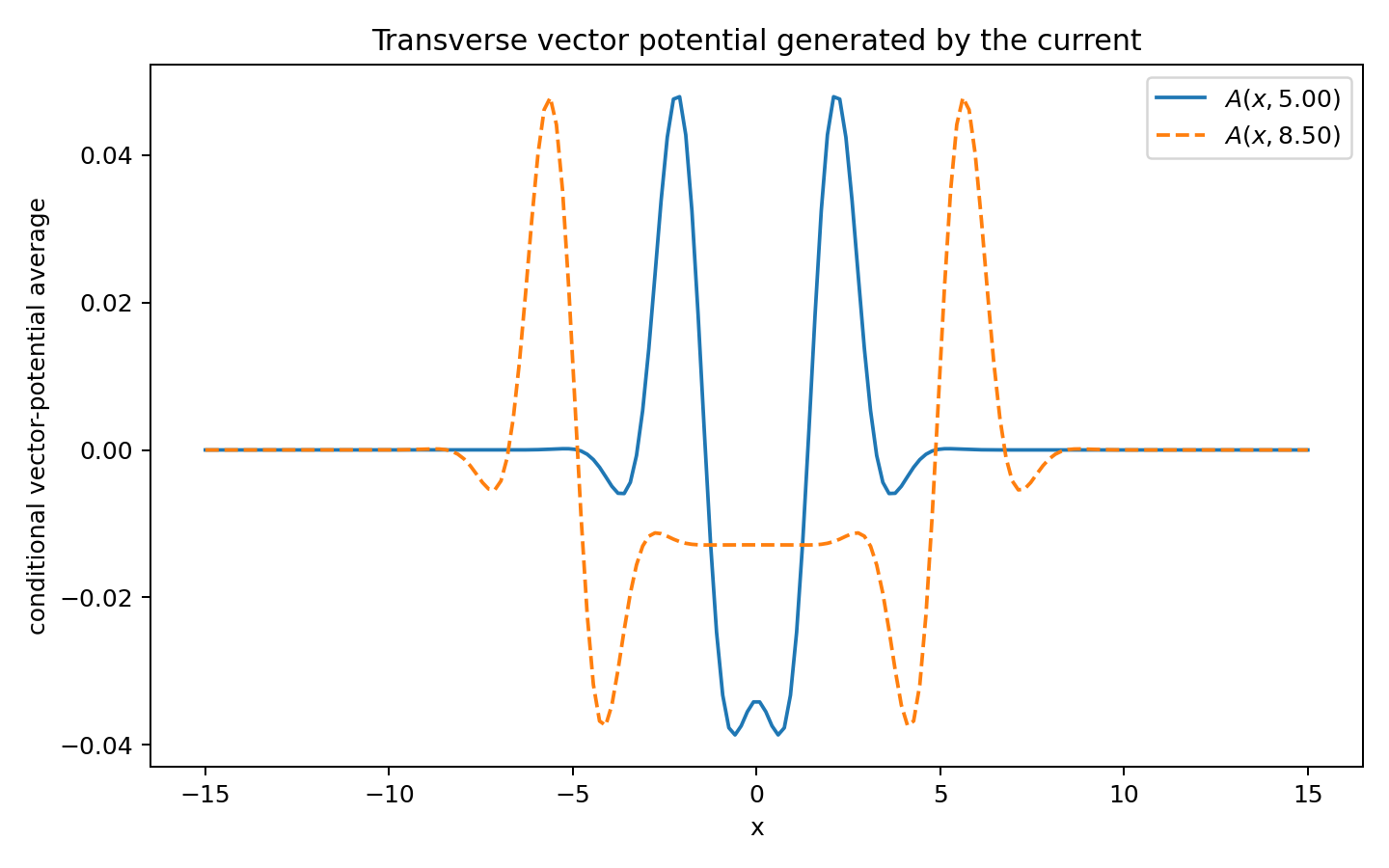}
\caption{Vector-potential coordinate \(\mathcal A(x,t)\) at two times. The current pulse generates outgoing transverse waves.}
\label{fig:em_vector_potential}
\end{figure}

Figure~\ref{fig:em_vector_potential} shows the vector-potential coordinate \(\mathcal A(x,t)\) at two different times. The current pulse generates outgoing waves moving away from the source region.

\begin{figure}[h]
\centering
\includegraphics[width=0.75\textwidth]{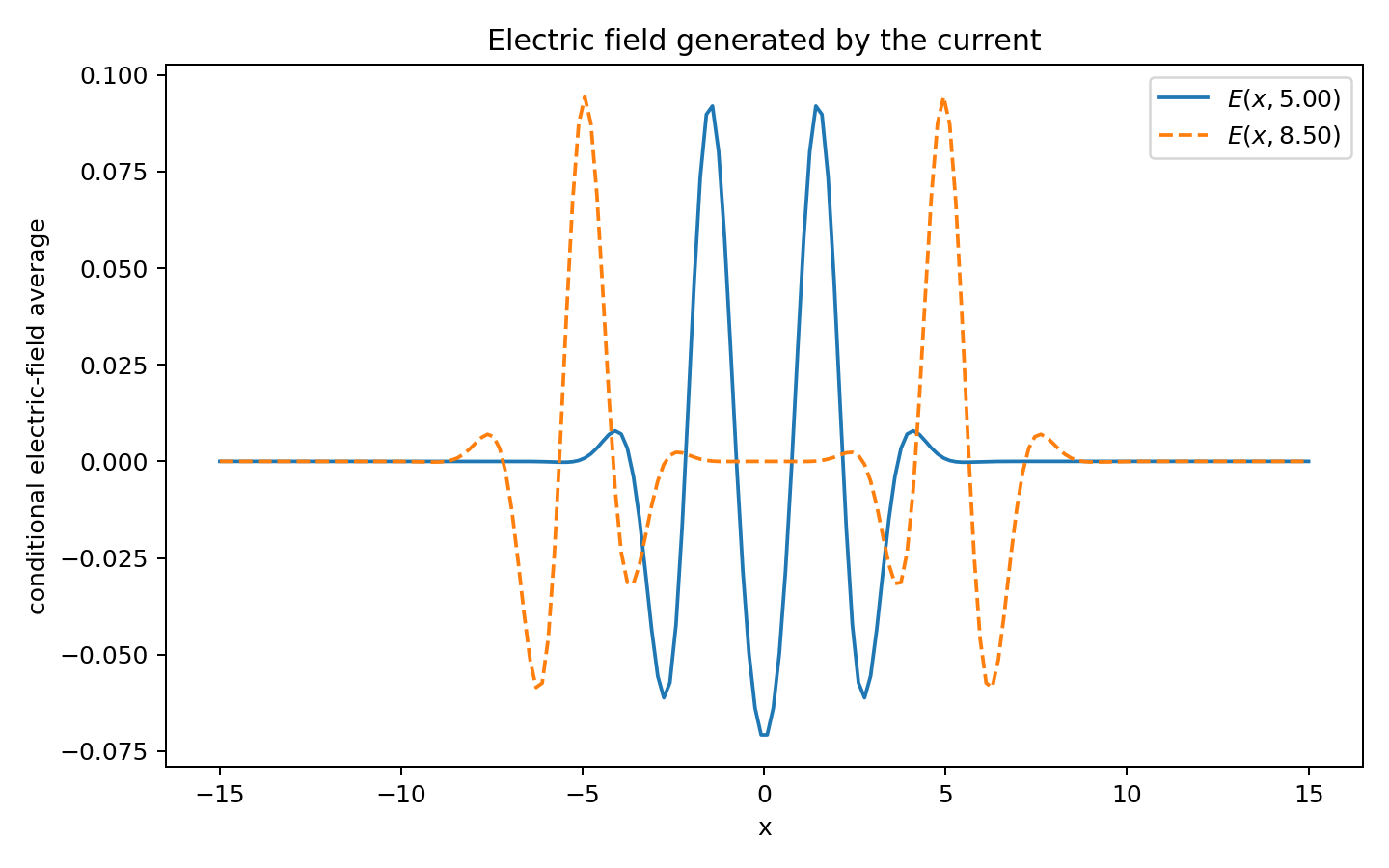}
\caption{Electric-field coordinate \(\mathcal E(x,t)=-\partial_t\mathcal A(x,t)\) at two times.}
\label{fig:em_electric_field}
\end{figure}

Figure~\ref{fig:em_electric_field} shows the corresponding electric-field coordinate
\[
\mathcal E(x,t)
=
-\partial_t\mathcal A(x,t)
\]
at the same two times.

\begin{figure}[h]
\centering
\includegraphics[width=0.75\textwidth]{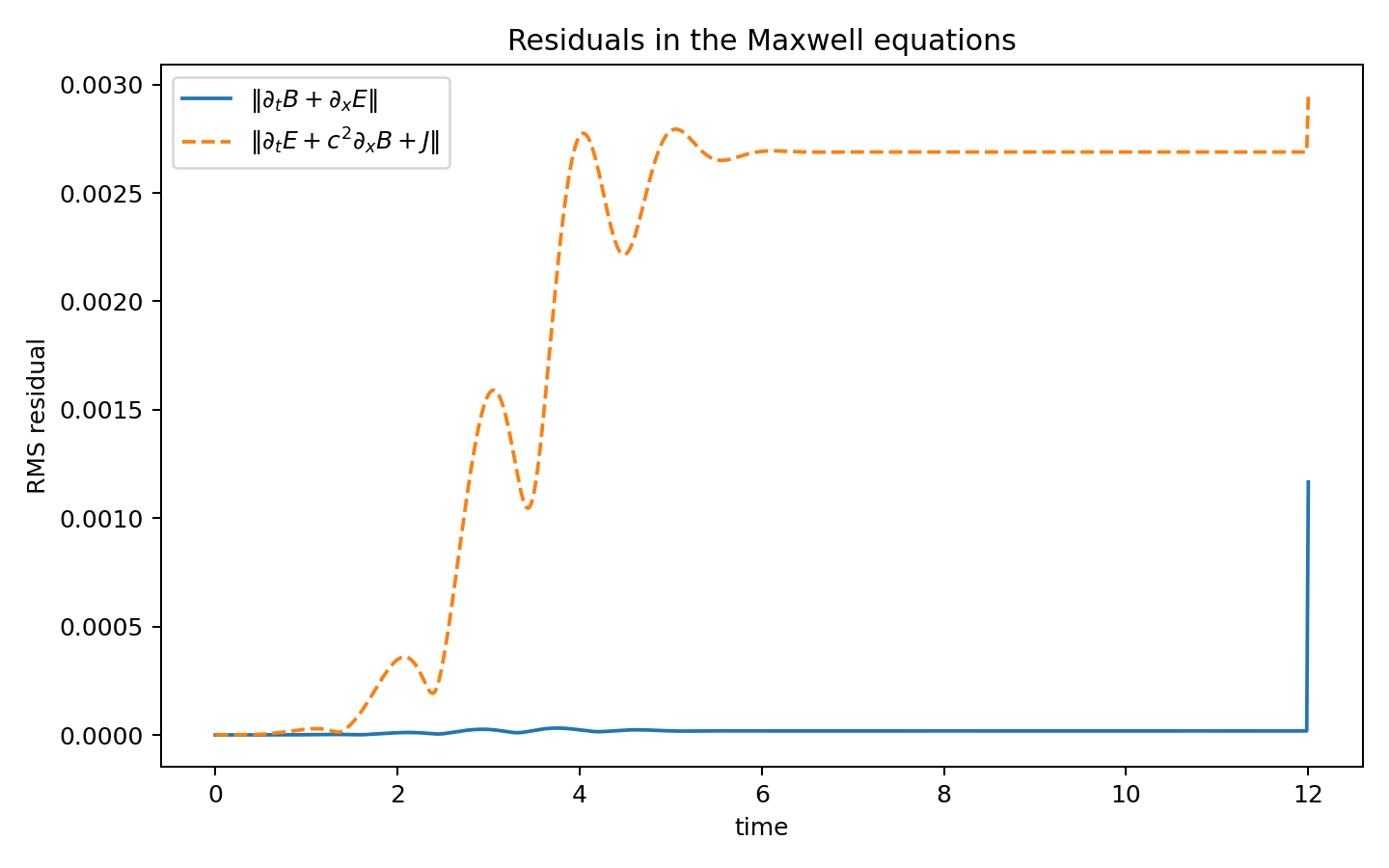}
\caption{RMS residuals in the one-dimensional Maxwell equations. The small residuals are due to numerical differentiation and finite-grid error.}
\label{fig:em_residuals}
\end{figure}

To verify the Maxwell equations directly, we computed the residuals
\[
R_F
=
\partial_t\mathcal B+\partial_x\mathcal E
\]
and
\[
R_A
=
\partial_t\mathcal E+c^2\partial_x\mathcal B+J .
\]
These are the residuals of Faraday's law and the transverse Ampere--Maxwell equation in the sign convention used above. The RMS residuals are shown in Figure~\ref{fig:em_residuals}. The maximum RMS residuals in the run were
\[
\max_t\|R_F(t)\|_{\mathrm{RMS}}
=
1.17\times 10^{-3},
\]
and
\[
\max_t\|R_A(t)\|_{\mathrm{RMS}}
=
2.94\times 10^{-3}.
\]
The nonzero values are due to the spatial and temporal discretization.

The simulation shows that the field coordinates selected by a smeared macroscopic current source obey the classical Maxwell equations. As in the scalar-field case, the field itself remains quantum in the functional Schr\"odinger representation. The current profile selects the relevant quotient-level field coordinates through the interaction term; the Gaussian functional used above is only a convenient representative of the corresponding source-selected equivalence class. Thus the classical electromagnetic field appears at the level of the field data accessible to the localized macroscopic source, not as a collapse of the full quantum field state. The geometric relation between these quotient coordinates and the tangential component of the quantum-field Schr\"odinger dynamics will be formulated below.

\section{Motion of a localized charge in the observed electromagnetic field}

We now simulate the motion of a localized test charge in the electromagnetic field obtained above. This step is important because classical field theory requires more than field coordinates satisfying Maxwell's equations. The field must also act on a localized charge by producing a definite force. This is precisely where localization of the field functional becomes essential.

Let
\[
\mathcal E(x,t)
=
-\partial_t \mathcal A(x,t),
\qquad
\mathcal B(x,t)
=
\partial_x \mathcal A(x,t),
\]
where \(\mathcal A(x,t)\) is the vector-potential coordinate on the quotient field manifold generated by the localized macroscopic current source. A localized charge with instantaneous position \(a(t)\) does not interact with point values of the field. Instead, its finite spatial profile selects the smeared field coordinates
\[
\mathcal E_W(a(t),t)
=
\int W_{a(t)}(x)\mathcal E(x,t)\,dx,
\]
and
\[
\mathcal B_W(a(t),t)
=
\int W_{a(t)}(x)\mathcal B(x,t)\,dx .
\]
In the narrow-packet approximation used in the simulation, these smeared coordinates are approximated by the field values sampled along the charge trajectory:
\[
\mathcal E_W(a(t),t)\approx \mathcal E(a(t),t),
\qquad
\mathcal B_W(a(t),t)\approx \mathcal B(a(t),t).
\]

For the one-dimensional transverse electromagnetic field considered above, we take
\[
\mathcal E=(0,\mathcal E_y,0),
\qquad
\mathcal B=(0,0,\mathcal B_z),
\]
with both components depending on \(x\) and \(t\). The nonrelativistic Lorentz-force equations for a charge moving in the \(xy\)-plane are then
\[
\dot x=v_x,
\qquad
\dot y=v_y,
\]
\[
\dot v_x
=
\frac{q}{m}v_y\mathcal B_z(x(t),t),
\]
and
\[
\dot v_y
=
\frac{q}{m}
\left(
\mathcal E_y(x(t),t)-v_x\mathcal B_z(x(t),t)
\right).
\]
Equivalently, in vector form,
\[
m\ddot a(t)
=
q\left(
\mathcal E_W(a(t),t)
+
\dot a(t)\times \mathcal B_W(a(t),t)
\right).
\]

These equations acquire the meaning of a definite classical force relative to
the smeared field observables selected by the localized charge. More precisely,
the charge profile \(W_{a(t)}\) selects the field observables entering the
Lorentz force,
\[
E_W(a(t),t),
\qquad
B_W(a(t),t).
\]
When, in an individual macroscopic realization, the corresponding quotient-level
values have dispersions
\[
\Delta E_W(a(t),t),
\qquad
\Delta B_W(a(t),t)
\]
small compared with the resolution of the macroscopic probe, they define a
definite classical force on the charge. Thus the emergence of classical
electrodynamics requires Maxwell equations for the selected field coordinates
together with effective definiteness of the probe-selected smeared observables
that act on charges.

\subsection{Numerical implementation}

We used the electromagnetic field obtained in the previous simulation and interpolated the field coordinates \(\mathcal E_y(x,t)\) and \(\mathcal B_z(x,t)\). A test charge was placed to the right of the source, initially at rest:
\[
x(0)=6,
\qquad
y(0)=0,
\qquad
v_x(0)=v_y(0)=0.
\]
The charge-to-mass ratio was chosen to be
\[
\frac{q}{m}=3.
\]
The charge then interacted with the outgoing electromagnetic pulse generated by the smeared current source.

The equations integrated numerically were
\[
\dot x=v_x,
\qquad
\dot y=v_y,
\]
\[
\dot v_x
=
\frac{q}{m}v_y\mathcal B_z(x(t),t),
\]
\[
\dot v_y
=
\frac{q}{m}
\left(
\mathcal E_y(x(t),t)-v_x\mathcal B_z(x(t),t)
\right).
\]
At each time step we recorded the charge position, velocity, field coordinates sampled along the trajectory, and Lorentz acceleration.

\subsection{Simulation results}

\begin{figure}[h]
\centering
\includegraphics[width=0.75\textwidth]{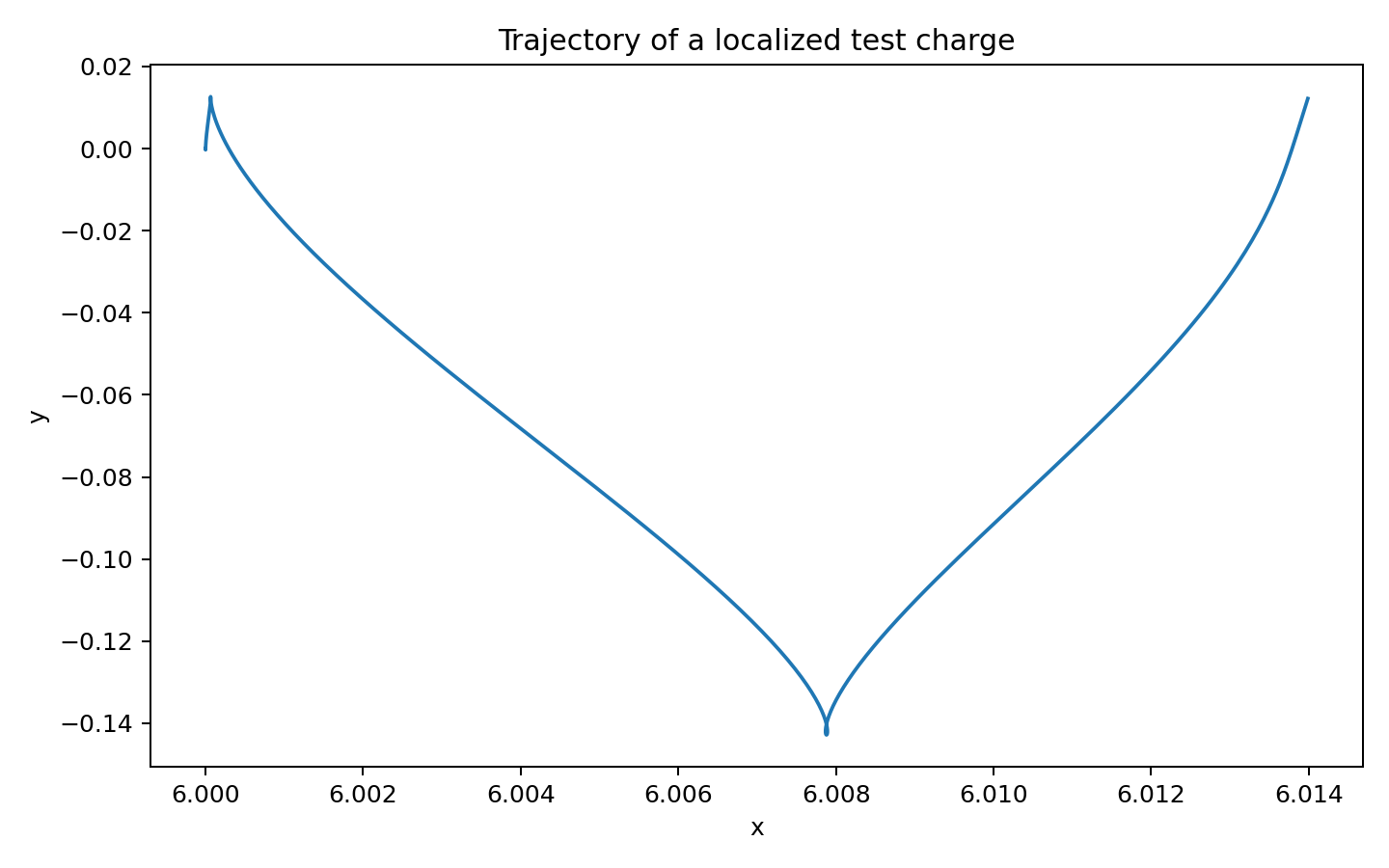}
\caption{Trajectory of a localized test charge in the electromagnetic field generated by the smeared current source. The charge is initially placed at \(x=6\), \(y=0\), and is accelerated when the outgoing field pulse reaches it.}
\label{fig:charge_trajectory_lorentz}
\end{figure}

The resulting trajectory is shown in Figure~\ref{fig:charge_trajectory_lorentz}. The charge is initially at rest. When the electromagnetic pulse reaches the charge, the transverse electric field accelerates it primarily in the \(y\)-direction, while the magnetic term produces a smaller \(x\)-deflection.

\begin{figure}[H]
\centering
\includegraphics[width=0.75\textwidth]{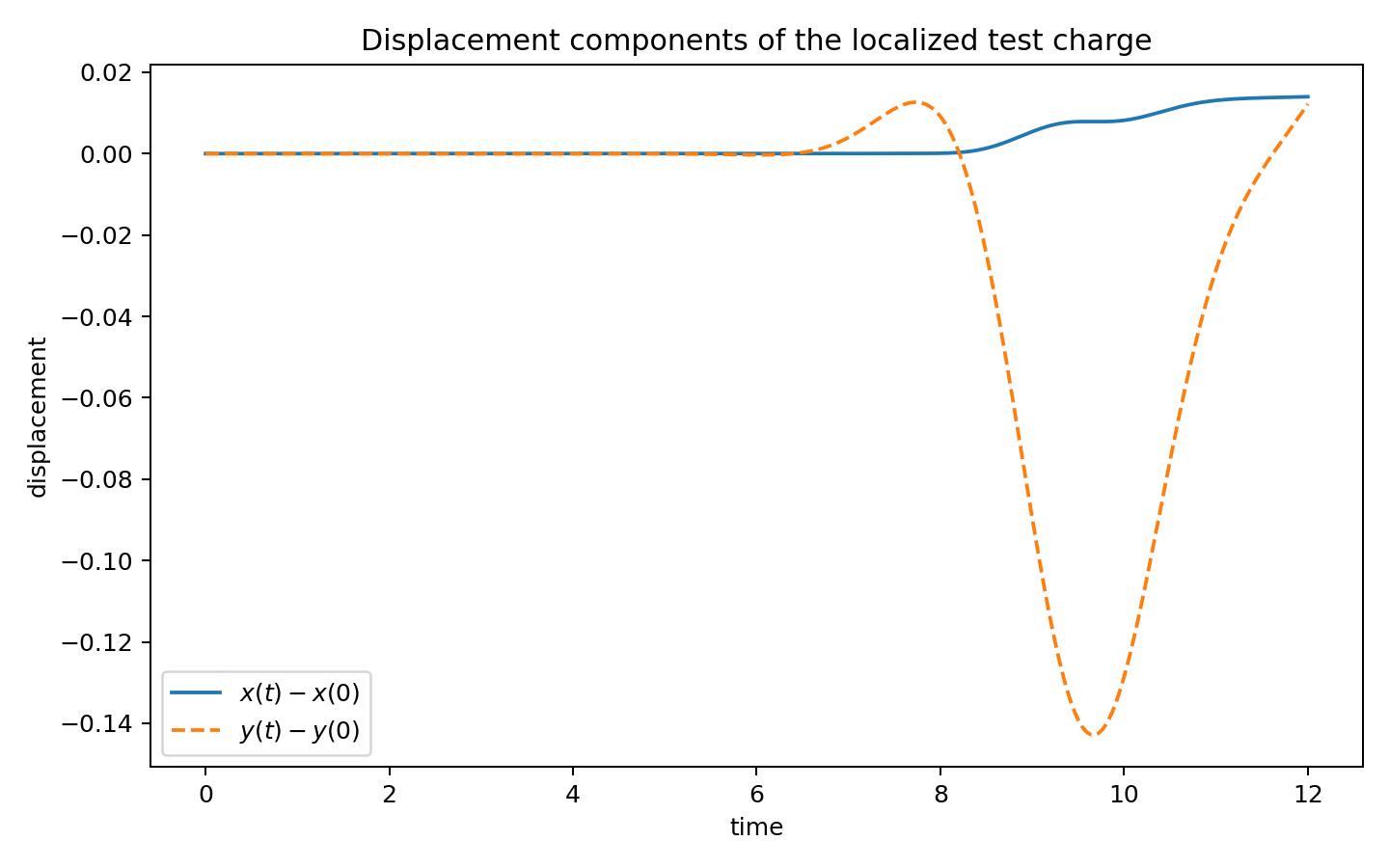}
\caption{Displacement components \(x(t)-x(0)\) and \(y(t)-y(0)\) of the localized test charge. Plotting the displacements relative to the initial position makes the smaller longitudinal motion visible together with the larger transverse response to the electromagnetic pulse.}
\label{fig:charge_position_components}
\end{figure}

Figure~\ref{fig:charge_position_components} shows the displacement components
\[
x(t)-x(0),
\qquad
y(t)-y(0).
\]
We plot displacements rather than absolute positions because the longitudinal motion is small compared with the initial value \(x(0)=6\). The transverse displacement records the main response of the charge to the electric-field pulse, while the smaller longitudinal displacement is produced by the magnetic term in the Lorentz force.

\begin{figure}[h]
\centering
\includegraphics[width=0.75\textwidth]{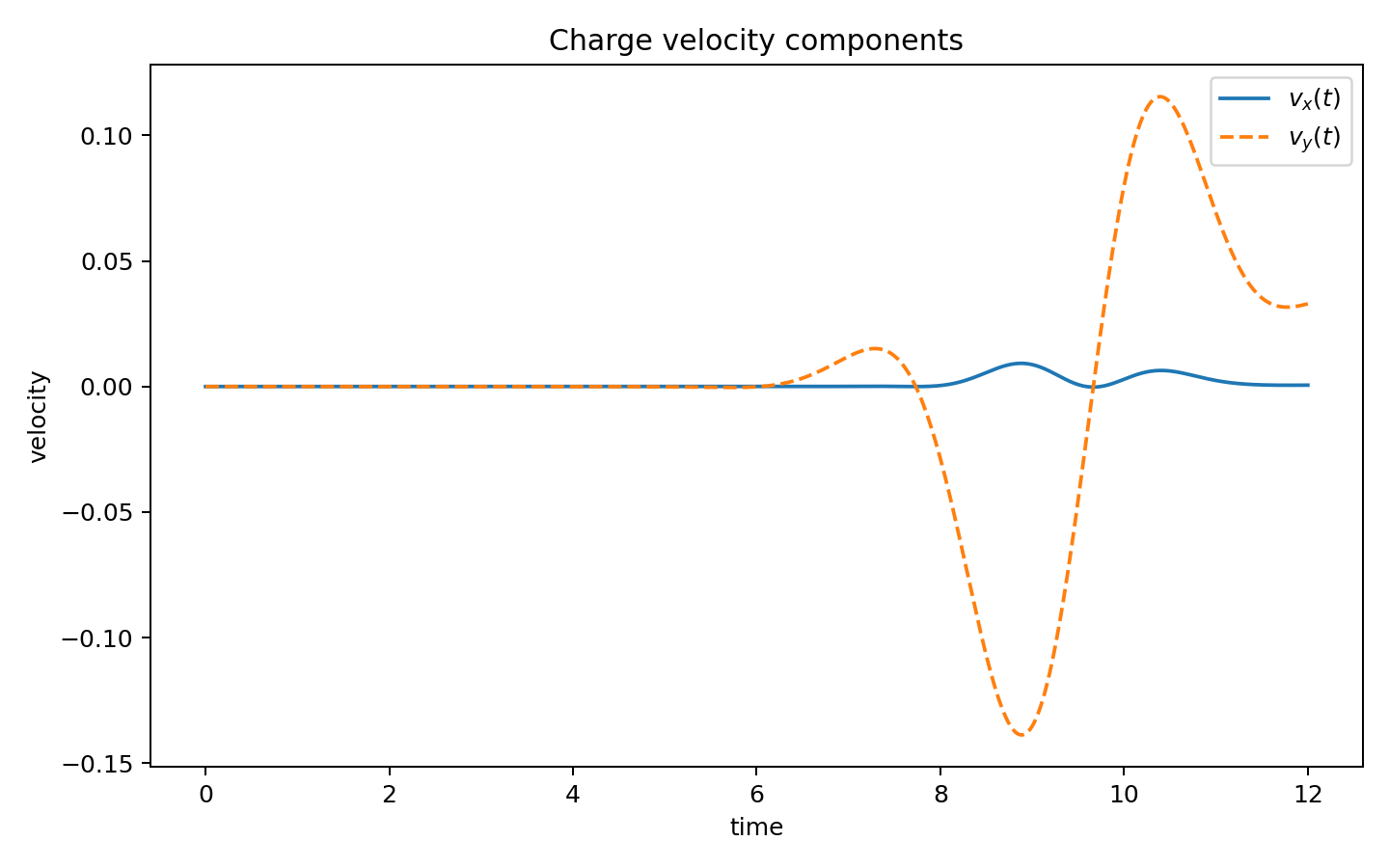}
\caption{Velocity components \(v_x(t)\) and \(v_y(t)\) of the localized test charge.}
\label{fig:charge_velocity_components}
\end{figure}

Figure~\ref{fig:charge_velocity_components} shows the velocity components. The main response is in \(v_y(t)\), as expected for a transverse electric field \(\mathcal E_y\). The smaller \(v_x(t)\) component is produced by the magnetic part of the Lorentz force.

\begin{figure}[h]
\centering
\includegraphics[width=0.75\textwidth]{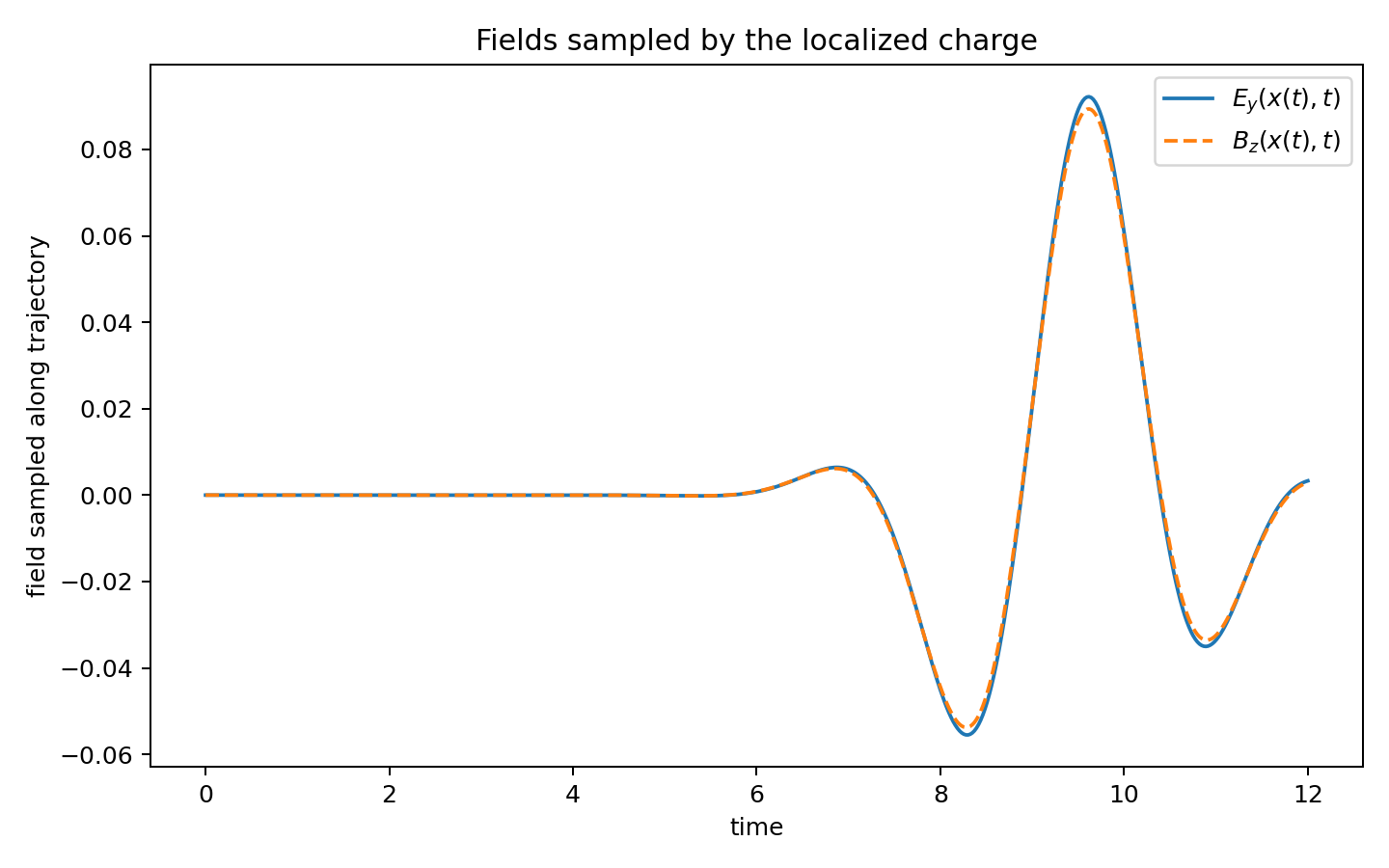}
\caption{Electromagnetic field coordinates sampled along the trajectory of the localized charge. These are the field coordinates entering the Lorentz-force equation.}
\label{fig:charge_sampled_fields}
\end{figure}

The field coordinates sampled along the charge trajectory are shown in Figure~\ref{fig:charge_sampled_fields}. These are the field coordinates entering the Lorentz-force equation. In the narrow-packet approximation, they approximate
\[
\mathcal E_W(a(t),t),
\qquad
\mathcal B_W(a(t),t),
\]
the field coordinates selected by the localized charge. The curves for
\[
\mathcal E_y(x(t),t)
\]
and
\[
\mathcal B_z(x(t),t)
\]
nearly coincide because the charge is located to the right of the source and is hit by the right-moving part of the transverse pulse. With the conventions
\[
\mathcal E_y=-\partial_t\mathcal A,
\qquad
\mathcal B_z=\partial_x\mathcal A,
\]
a right-moving wave
\[
\mathcal A=f(x-ct)
\]
satisfies
\[
\mathcal E_y=c\mathcal B_z.
\]
Since the simulation uses \(c=1\), the two sampled field coordinates agree up to numerical error.

\begin{figure}[h]
\centering
\includegraphics[width=0.75\textwidth]{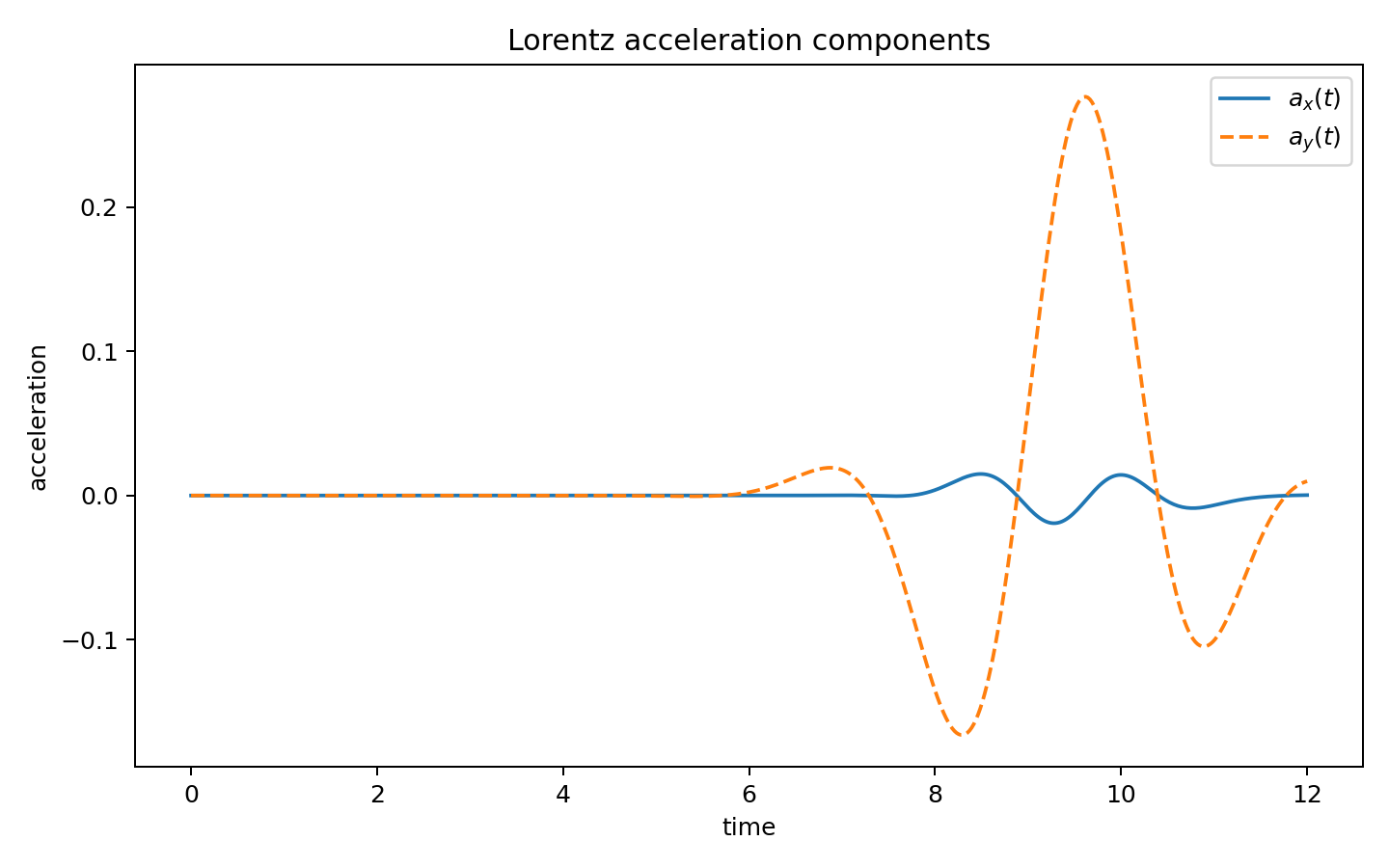}
\caption{Lorentz acceleration components of the localized test charge. The transverse acceleration is produced mainly by the electric field, while the smaller longitudinal acceleration is produced by the magnetic term.}
\label{fig:charge_lorentz_acceleration}
\end{figure}

Figure~\ref{fig:charge_lorentz_acceleration} shows the acceleration components computed from the Lorentz force:
\[
a_x(t)
=
\frac{q}{m}v_y(t)\mathcal B_z(x(t),t),
\]
\[
a_y(t)
=
\frac{q}{m}
\left(
\mathcal E_y(x(t),t)-v_x(t)\mathcal B_z(x(t),t)
\right).
\]
The acceleration is nonzero only when the localized charge encounters the electromagnetic pulse.

The final position and velocity were
\[
x(12)=6.013989,
\qquad
y(12)=0.012188,
\]
and
\[
v_x(12)=0.000545,
\qquad
v_y(12)=0.032964.
\]
The maximum sampled field strengths along the trajectory were
\[
\max_t |\mathcal E_y(x(t),t)|
=
9.22\times 10^{-2},
\]
and
\[
\max_t |\mathcal B_z(x(t),t)|
=
8.94\times 10^{-2}.
\]
The maximum Lorentz acceleration was
\[
\max_t |\ddot a(t)|
=
2.77\times 10^{-1}.
\]

This simulation completes the operational picture. The previous electromagnetic
simulation showed that the field coordinates selected by a macroscopic current
source satisfy Maxwell's equations. The present simulation shows that the field
coordinates sampled by a localized charge enter the Lorentz-force law for that
charge. Thus the classical electromagnetic field appears operationally as the
quotient-level field data selected by localized macroscopic sources and probes.

\subsection{Equivalence classes of localized field functionals}

The preceding simulations used Gaussian field functionals as convenient representatives of localized field states. We now explain why Gaussian representatives are sufficient for defining, locally, the metric on the quotient manifold of equivalence classes induced by the Fubini--Study metric. In the finite-dimensional field approximation this can be verified directly.

Let the discretized field configuration be
\[
q=(q_1,\ldots,q_N)\in\mathbb R^N,
\]
and let the field state be a normalized wave function
\[
\Psi(q)\in L^2(\mathbb R^N).
\]
A macroscopic probe has finite resolution and cannot distinguish all field functionals. It only accesses coarse-grained field coordinates. For simplicity, first take these coordinates to be the lattice field values \(q_j\) themselves. More general coordinates are obtained by replacing \(q\) with \(Wq\).

Fix a positive definite covariance matrix \(\Sigma\), representing the resolution of the macroscopic probe. We define the localized equivalence class centered at \(Q\in\mathbb R^N\) by
\[
[Q]_\Sigma
=
\left\{
\Psi\in L^2(\mathbb R^N):
\|\Psi\|=1,\quad
\langle q\rangle_\Psi=Q,\quad
\operatorname{Cov}_\Psi(q)\leq \Sigma
\right\}.
\]
States in this class may have different functional forms. They need not be Gaussian. They are equivalent because they have the same field center \(Q\), within the prescribed resolution, and their fluctuations in the accessible field variables are below the resolution scale.

The distance between two equivalence classes is defined by
\[
\rho([Q]_\Sigma,[Q']_\Sigma)
=
\inf_{\Psi\in[Q]_\Sigma,\ \Phi\in[Q']_\Sigma}
\rho_{\mathrm{FS}}(\Psi,\Phi),
\]
where \(\rho_{\mathrm{FS}}\) is the Fubini--Study distance. We show that, to leading order for nearby classes, this distance is determined by Gaussian representatives.

Let \(\Psi(q)=\sqrt{\rho(q)}\) be a real localized representative with mean \(Q\), and consider the translated state
\[
\Psi_{dQ}(q)=\Psi(q-dQ).
\]
For small \(dQ\), the induced Fubini--Study line element is
\[
ds^2
=
\int_{\mathbb R^N}
\left|
\nabla\Psi(q)\cdot dQ
\right|^2 dq .
\]
Since \(\Psi=\sqrt{\rho}\), this can be written as
\[
ds^2
=
\frac14\, dQ^T I_\rho\, dQ,
\]
where \(I_\rho\) is the Fisher-information matrix
\[
(I_\rho)_{ij}
=
\int_{\mathbb R^N}
\frac{\partial_i\rho(q)\partial_j\rho(q)}{\rho(q)}\,dq .
\]
The multivariate Cram\'er--Rao inequality gives
\[
I_\rho
\geq
\operatorname{Cov}_\rho(q)^{-1}.
\]
Since the state belongs to the localized class, we have
\[
\operatorname{Cov}_\rho(q)\leq \Sigma.
\]
Therefore
\[
I_\rho\geq \Sigma^{-1},
\]
and hence
\[
ds^2
\geq
\frac14\, dQ^T\Sigma^{-1}dQ.
\]

Equality is achieved by the Gaussian representative
\[
G_{Q,\Sigma}(q)
=
C
\exp\left[
-\frac14(q-Q)^T\Sigma^{-1}(q-Q)
\right],
\]
whose probability density has covariance \(\Sigma\). Thus, to leading order,
\[
\rho([Q]_\Sigma,[Q+dQ]_\Sigma)^2
=
\frac14\, dQ^T\Sigma^{-1}dQ
+
o(|dQ|^2).
\]

The Gaussian representative therefore determines the induced metric on the space of localized equivalence classes, but the equivalence class itself contains many non-Gaussian field functionals.
The geometry is assigned to the equivalence classes
\([Q]_\Sigma\) themselves, rather than to the chosen Gaussian representatives.
The operational interpretation of these classes in terms of field data
accessible to macroscopic probes was illustrated in the simulations above and
will be justified dynamically below.

\section{Tangential quantum-field dynamics and classical field theory}
\label{sec:tangent}

We now formulate the geometric meaning of the preceding simulations. The quantum field is not assumed to become classical. It remains represented by a state in the field Hilbert space. Classical fields are represented by localized equivalence classes of quantum field states, while the corresponding classical field variables appear as coordinates labeling these classes. Classical field equations arise as the tangential component of the quantum-field Schr\"odinger dynamics on the corresponding localized quotient manifold.

In a finite-dimensional field approximation, let
\[
q=(q_1,\ldots,q_N)\in\mathbb R^N
\]
denote the field-configuration variables, and let
\[
\widehat p_j
=
-i\hbar\frac{\partial}{\partial q_j}
\]
be the conjugate momentum operators. A localized phase-space equivalence class is labeled by coordinates
\[
(Q,P)
=
(Q_1,\ldots,Q_N,P_1,\ldots,P_N).
\]
The class \([Q,P]_\Sigma\) consists of localized field states whose field-position coordinates are \(Q_j\), whose conjugate-momentum coordinates are \(P_j\), and whose field-configuration spread is bounded by the resolution scale \(\Sigma\). Thus the quotient phase-space manifold is
\[
\mathcal M_{\mathrm{field}}^\Sigma
=
\{[Q,P]_\Sigma:(Q,P)\in\mathbb R^{2N}\}.
\]
The coordinates \((Q,P)\) are the finite-dimensional classical field variables.

Gaussian representatives provide a convenient way to label nearby localized
equivalence classes. For fixed covariance \(\Sigma\), we use
\[
G_{Q,P,\Sigma}(q)
=
C
\exp\left[
-\frac14(q-Q)^T\Sigma^{-1}(q-Q)
+
\frac{i}{\hbar}P\cdot q
\right].
\]
Here \(Q\) and \(P\) are quotient coordinates, while \(\Sigma\) is fixed and
represents the localization width used in defining the localized classes. The
Gaussian state is not part of the physical definition of the classical field.
The physical object is the localized equivalence class \([Q,P]_\Sigma\). The
Gaussian representative is used only because it makes the induced metric and
symplectic form explicit.

For a normalized field state \(\Psi\), the quantum action is
\[
S_Q[\Psi]
=
\int
\left(
i\hbar\langle \Psi,\dot\Psi\rangle
-
\langle \Psi,\widehat H\Psi\rangle
\right)dt .
\]
This action is naturally defined on projective Hilbert space: a change of phase
changes the first term only by a total derivative. To restrict the action to
paths in the localized quotient sector, we choose smooth representatives
\[
(Q,P)\mapsto \Psi_{Q,P}.
\]
This gives
\[
S_Q\big|_{\mathcal M^\Sigma_{\mathrm{field}}}
=
\int
\left[
i\hbar
\left\langle
\Psi_{Q,P},
\frac{d}{dt}\Psi_{Q,P}
\right\rangle
-
\left\langle
\Psi_{Q,P},
\widehat H\Psi_{Q,P}
\right\rangle
\right]dt .
\]
Different choices of smooth representatives change the first term by an exact
differential and therefore do not change the equations of motion on the
quotient manifold.

The pullback of the projective symplectic potential to the localized quotient sector has the canonical form
\[
i\hbar
\langle
\Psi_{Q,P},
d\Psi_{Q,P}
\rangle
=
\sum_{j=1}^N P_j\,dQ_j
+
dF
+
O(\Sigma),
\]
where \(dF\) is an exact differential and \(O(\Sigma)\) denotes corrections controlled by the localization width. Hence the induced symplectic form is
\[
\omega_{\mathrm{FS}}\big|_{\mathcal M^\Sigma_{\mathrm{field}}}
=
\sum_{j=1}^N dP_j\wedge dQ_j
+
O(\Sigma).
\]
The Hamiltonian induced on the quotient manifold is
\[
H_{\mathrm{cl}}(Q,P)
=
\left\langle
\Psi_{Q,P},
\widehat H\Psi_{Q,P}
\right\rangle .
\]
Therefore, up to irrelevant boundary terms and fixed-width corrections, the restricted action takes the classical Hamiltonian form
\[
S_Q\big|_{\mathcal M^\Sigma_{\mathrm{field}}}
=
\int
\left(
\sum_{j=1}^N P_j\dot Q_j
-
H_{\mathrm{cl}}(Q,P)
\right)dt .
\]
Varying this action gives
\[
\dot Q_j
=
\frac{\partial H_{\mathrm{cl}}}{\partial P_j},
\qquad
\dot P_j
=
-
\frac{\partial H_{\mathrm{cl}}}{\partial Q_j}.
\]
Thus the classical field equations are the Hamilton equations induced by the quantum action restricted to the localized quotient manifold.

Equivalently, this is the tangent projection of the Schr\"odinger velocity onto the localized quotient sector. If
\[
V_\Psi
=
-\frac{i}{\hbar}\widehat H\Psi
\]
is the Hilbert-space Schr\"odinger velocity, then its tangent component along \(\mathcal M^\Sigma_{\mathrm{field}}\) induces the above Hamiltonian flow on the quotient coordinates \((Q,P)\). In this sense the classical field dynamics is the tangential part of the quantum-field Schr\"odinger dynamics.

For quadratic field Hamiltonians with linear source terms, the expectation value \(H_{\mathrm{cl}}(Q,P)\) differs from the usual classical Hamiltonian only by terms depending on the fixed localization width \(\Sigma\). These terms are independent of \(Q\) and \(P\), and therefore do not affect the Hamilton equations. Consequently, for quadratic field Hamiltonians, the tangential equations close exactly on the localized quotient coordinates. For non-quadratic Hamiltonians, the same construction gives the leading tangential dynamics in the localized sector, with corrections controlled by the spread of the field functional.

\subsection{Scalar field}

For a discretized scalar field, take
\[
\widehat H
=
\frac12\sum_{j=1}^N \widehat p_j^2
+
\frac12\sum_{j,k=1}^N q_jK_{jk}q_k
-
\sum_{j=1}^N J_jq_j,
\]
where \(K\) is the discrete Klein--Gordon operator. The induced Hamiltonian on the localized quotient manifold is
\[
H_{\mathrm{cl}}(Q,P)
=
\frac12\sum_{j=1}^N P_j^2
+
\frac12\sum_{j,k=1}^N Q_jK_{jk}Q_k
-
\sum_{j=1}^N J_jQ_j
+
C_\Sigma,
\]
where \(C_\Sigma\) depends on the fixed localization width but is independent of \(Q\) and \(P\). Therefore
\[
\dot Q_j=P_j,
\qquad
\dot P_j
=
-\sum_{k=1}^N K_{jk}Q_k+J_j,
\]
and hence
\[
\ddot Q_j+\sum_{k=1}^N K_{jk}Q_k=J_j.
\]
In the continuum limit this becomes
\[
\partial_t^2\Phi(x,t)-\Delta\Phi(x,t)+m^2\Phi(x,t)
=
J(x,t),
\]
which is the sourced Klein--Gordon equation.

Thus the classical scalar field \(\Phi\) is the coordinate on the localized quotient manifold, and the Klein--Gordon equation is the tangential component of the quantum-field Schr\"odinger dynamics on this manifold.

\subsection{Electromagnetic field}

For the transverse electromagnetic field, take a finite-dimensional approximation with vector-potential variables \(A_j\) and conjugate momenta \(\Pi_j\). The quantum Hamiltonian has the form
\[
\widehat H
=
\frac12\sum_j \widehat\Pi_j^2
+
\frac12 c^2\sum_{j,k} A_jL_{jk}A_k
-
\sum_j J_jA_j,
\]
where \(L\) is the discrete spatial derivative operator associated with the magnetic-field energy. The localized quotient coordinates are denoted by
\[
\mathcal A_j,
\qquad
\mathcal P_j.
\]
The induced Hamiltonian is
\[
H_{\mathrm{cl}}(\mathcal A,\mathcal P)
=
\frac12\sum_j \mathcal P_j^2
+
\frac12 c^2\sum_{j,k}\mathcal A_jL_{jk}\mathcal A_k
-
\sum_j J_j\mathcal A_j
+
C_\Sigma,
\]
where \(C_\Sigma\) is independent of \(\mathcal A\) and \(\mathcal P\). Therefore
\[
\dot{\mathcal A}_j=\mathcal P_j,
\]
and
\[
\dot{\mathcal P}_j
=
-c^2\sum_k L_{jk}\mathcal A_k+J_j.
\]
Equivalently,
\[
\ddot{\mathcal A}_j
+
c^2\sum_k L_{jk}\mathcal A_k
=
J_j.
\]
In the continuum limit, this gives the transverse wave equation
\[
\partial_t^2\mathcal A-c^2\Delta\mathcal A=J.
\]
Defining
\[
\mathcal E=-\partial_t\mathcal A,
\qquad
\mathcal B=\nabla\times\mathcal A,
\]
we obtain Maxwell's equations for the field coordinates:
\[
\partial_t\mathcal B=-\nabla\times\mathcal E,
\]
and
\[
\partial_t\mathcal E
=
c^2\nabla\times\mathcal B
-
J,
\]
with the sign convention determined by the interaction term in the Hamiltonian.

\subsection{Limitations of coherent-state and Ehrenfest arguments}

Coherent states are often used as a bridge from quantum fields to classical fields because their field coordinates may satisfy classical-looking equations and their fluctuations can be small. In the present framework, however, coherent states are not the origin of classicality, nor are they the essential representatives of classical fields.
The physical object is not a coherent state, nor any particular Gaussian representative, but the localized equivalence class of field states. Gaussian representatives, rather than coherent states as such, are convenient for computing the induced Fubini--Study geometry on the quotient manifold.

A classical field is represented by coordinates on the quotient manifold of equivalence classes, associated with the field observables accessible to macroscopic probes. Non-Gaussian localized functionals in the same equivalence class correspond to the same point on the quotient manifold and therefore represent the same classical field. Thus classicality is not the selection of coherent states as special physical states, but the restriction of macroscopic observations to localized quotient sectors of quantum field state space.

This also explains why an argument based only on Ehrenfest equations is not sufficient. In quantum mechanics, a sufficiently complete system of equations for expectation values is essentially another form of the Schr\"odinger dynamics itself. Such equations do not by themselves select localized states, classical trajectories, or classical fields; they describe the evolution of expectation values. The quotient-and-localization structure is therefore the essential ingredient missing from a purely Ehrenfest-based account of classicality: it identifies the localized quotient sector on which the tangential component of the Schr\"odinger dynamics becomes classical field dynamics.

We summarize the identification as
\[
\boxed{
\begin{aligned}
\mathsf{Classical\ field\ variables}
&=
\mathsf{coordinates\ on\ localized\ equivalence\ classes} \\
&\quad
\mathsf{of\ quantum\ field\ states},
\end{aligned}
}
\]
and
\[
\boxed{
\begin{aligned}
\mathsf{Classical\ field\ equations}
&=
\mathsf{tangential\ components\ of\ the\ quantum\text{-}field} \\
&\quad
\mbox{\sffamily Schr\"odinger dynamics}.
\end{aligned}
}
\]

\subsection{Compatibility with random-matrix localization}

We finally explain why the random-matrix stabilization mechanism {\bf (RM)} of \cite{KryukovPHLA,KryukovPhysicsA,KryukovArx,KryukovCompArx} is compatible with the field interpretation presented above. The total dynamics contains two distinct effects. The first is the smooth Hamiltonian motion of a localized charge in the observed electromagnetic field. The second is the rapid environmental stabilization already present in the particle case and represented by {\bf (RM)}. The electromagnetic field provides an additional interaction acting on the charge; it does not replace the environmental interaction responsible for {\bf (RM)} stabilization.

Schematically, the total Hamiltonian has the form
\[
\widehat H
=
\widehat H_{\rm field}
+
\widehat H_{\rm ch}
+
\widehat H_{\rm int}
+
\widehat H_{\rm RM}(t),
\]
where \(\widehat H_{\rm int}\) is the ordinary charge-field interaction producing the Lorentz force, while \(\widehat H_{\rm RM}(t)\) represents the effective particle-environment interaction responsible for localization near the classical particle manifold.

The required compatibility condition is not exact commutation with the full Hamiltonian. Rather, the random-matrix interaction must act, to the required approximation, on the particle/environment localization degrees of freedom without disturbing the field observables during the short environmental stabilization step. Thus, for the relevant smeared field observables,
\[
[\widehat H_{\rm RM}(t),\widehat{\mathcal O}_{\rm field}]
\approx 0,
\]
where \(\widehat{\mathcal O}_{\rm field}\) denotes observables such as
\[
\phi_W(a),
\qquad
A_W(a),
\qquad
E_W(a),
\qquad
B_W(a).
\]
This expresses the assumption that the environment records the particle position without directly measuring or randomizing the field variables.

Let \(\Gamma_{\rm env}\) be the rate of environmental interactions responsible for {\bf (RM)} stabilization. Then
\[
\tau_{\rm env}
=
\Gamma_{\rm env}^{-1}
\]
is the characteristic time between localization events. The electromagnetic field produces, during this time, a displacement of order
\[
\Delta a_{\rm EM}
\sim
v\,\tau_{\rm env}
+
\frac12
\frac{Q}{M}
\left(E+vB\right)
\tau_{\rm env}^2,
\]
where \(M\) and \(Q\) are the mass and charge of the macroscopic body. The separation condition is
\[
\Delta a_{\rm EM}
\ll
\sigma,
\]
where \(\sigma\) is the spatial width of the localized particle state. Equivalently,
\[
\Gamma_{\rm env}
\gg
\max
\left\{
\frac{v}{\sigma},
\sqrt{
\frac{(Q/M)(E+vB)}{2\sigma}
}
\right\}.
\]

For a macroscopic body in air, using the gas-collision estimates of \cite{KryukovPHLA}, the environmental collision rate is approximately
\[
\Gamma_{\rm env}
\sim
n_{\rm air}\bar v\,\pi R^2,
\]
where \(R\) is the body radius, \(n_{\rm air}\) is the density of air molecules, and \(\bar v\) is their thermal velocity. Taking
\[
n_{\rm air}
\sim
2.5\times 10^{25}\,{\rm m}^{-3},
\qquad
\bar v
\sim
500\,{\rm m/s},
\qquad
R
=
10^{-6}\,{\rm m},
\]
we find
\[
\Gamma_{\rm env}
\sim
4\times 10^{16}\,{\rm s}^{-1},
\]
and hence
\[
\tau_{\rm env}
\sim
2.5\times 10^{-17}\,{\rm s}.
\]
This is for a micron-sized particle; larger macroscopic bodies have even larger environmental interaction rates.

Now take a relatively large charge-to-mass ratio and a strong electric field,
\[
\frac{Q}{M}
=
1\,{\rm C/kg},
\qquad
E
=
10^6\,{\rm V/m}.
\]
Then
\[
a_{\rm EM}
=
\frac{Q}{M}E
=
10^6\,{\rm m/s^2}.
\]
The acceleration contribution to the displacement between environmental interactions is
\[
\frac12 a_{\rm EM}\tau_{\rm env}^2
\sim
\frac12
(10^6)
(2.5\times 10^{-17})^2
\sim
3\times 10^{-28}\,{\rm m}.
\]
Even the displacement due to an existing velocity is negligible. For example, if
\[
v
=
1\,{\rm m/s},
\]
then
\[
v\tau_{\rm env}
\sim
2.5\times 10^{-17}\,{\rm m}.
\]
Both values are vastly smaller than any macroscopic localization width, for example
\[
\sigma
=
10^{-7}\,{\rm m}.
\]
Thus, in a natural environment, the {\bf (RM)} stabilization acts on a time scale enormously shorter than the Lorentz motion caused by ordinary electromagnetic fields. 

 In air, the dominant environmental interactions are collisions with air molecules. In vacuum, the relevant natural environment is radiation: ambient thermal photons and photons emitted by the macroscopic body itself. We estimate this rate for a micron-sized body at room temperature.
The number density of blackbody photons at temperature \(T\) is
\[
n_\gamma(T)
=
\frac{2\zeta(3)}{\pi^2}
\left(
\frac{k_B T}{\hbar c}
\right)^3 .
\]
For \(T=300\,{\rm K}\), this gives
\[
n_\gamma(300\,{\rm K})
\sim
5\times 10^{14}\,{\rm m}^{-3}.
\]
The incident photon rate on a body of radius \(R\) is approximately
\[
\Gamma_{\gamma,{\rm in}}
\sim
\frac14 c\,n_\gamma\,\pi R^2 .
\]
For
\[
R=10^{-6}\,{\rm m},
\]
we obtain
\[
\Gamma_{\gamma,{\rm in}}
\sim
\frac14
(3\times 10^8)
(5\times 10^{14})
\pi(10^{-6})^2
\sim
10^{11}\,{\rm s}^{-1}.
\]

The body also emits thermal photons. Its radiated power is approximately
\[
P_{\rm rad}
\sim
4\pi R^2 \epsilon \sigma_{\rm SB} T^4,
\]
where \(\epsilon\) is the emissivity and \(\sigma_{\rm SB}\) is the Stefan--Boltzmann constant. The mean energy of a blackbody photon is of order
\[
\langle E_\gamma\rangle
\sim
2.7 k_B T.
\]
Thus the photon-emission rate is
\[
\Gamma_{\gamma,{\rm em}}
\sim
\frac{4\pi R^2 \epsilon \sigma_{\rm SB} T^4}{2.7 k_B T}.
\]
For \(R=10^{-6}\,{\rm m}\), \(T=300\,{\rm K}\), and \(\epsilon\) of order one,
\[
\Gamma_{\gamma,{\rm em}}
\sim
5\times 10^{11}\,{\rm s}^{-1}.
\]
Thus even in vacuum, a micron-sized room-temperature body interacts with the radiation environment at a rate of order
\[
\Gamma_{\rm env}
\sim
10^{11}\text{--}10^{12}\,{\rm s}^{-1}.
\]
The corresponding environmental time scale is
\[
\tau_{\rm env}
=
\Gamma_{\rm env}^{-1}
\sim
10^{-12}\text{--}10^{-11}\,{\rm s}.
\]

We now compare this time scale with the Lorentz motion caused by an external electromagnetic field. 
For an upper-bound estimate, we use the relatively large charge-to-mass ratio and strong electric field considered above,
\[
\frac{Q}{M}=1\,{\rm C/kg},
\qquad
E=10^6\,{\rm V/m}.
\]
Then
\[
a_{\rm EM}
=
\frac{Q}{M}E
=
10^6\,{\rm m/s^2}.
\]
For the conservative value
\[
\tau_{\rm env}=10^{-11}\,{\rm s},
\]
the acceleration contribution to the displacement between environmental interactions is
\[
\frac12 a_{\rm EM}\tau_{\rm env}^2
\sim
\frac12
(10^6)(10^{-11})^2
=
5\times 10^{-17}\,{\rm m}.
\]
The displacement due to an already existing velocity is also tiny. For example, for
\[
v=1\,{\rm m/s},
\]
one has
\[
v\tau_{\rm env}
\sim
10^{-11}\,{\rm m}.
\]
Both estimates are far below any macroscopic localization scale. Even if the localization width is taken to be as small as
\[
\sigma=10^{-7}\,{\rm m},
\]
we have
\[
\Delta a_{\rm EM}
\ll
\sigma.
\]

The evolution can therefore be separated into rapid environmental localization steps and smooth electromagnetic motion between such steps. Over one environmental time interval, one may write schematically
\[
U(\tau_{\rm env})
\approx
U_{\rm EM}(\tau_{\rm env})\,
U_{\rm RM}(\tau_{\rm env}),
\]
where
\[
U_{\rm EM}(\tau_{\rm env})
=
\exp
\left[
-\frac{i}{\hbar}
\left(
\widehat H_{\rm field}
+
\widehat H_{\rm ch}
+
\widehat H_{\rm int}
\right)
\tau_{\rm env}
\right],
\]
and
\[
U_{\rm RM}(\tau_{\rm env})
=
\exp
\left[
-\frac{i}{\hbar}
\widehat H_{\rm RM}
\tau_{\rm env}
\right].
\]
Here \(U_{\rm RM}\) represents the stochastic environmental interaction supplied by air collisions, thermal radiation, or any other environmental process capable of producing the random-matrix stabilization described in \cite{KryukovPHLA,KryukovPhysicsA,KryukovArx,KryukovCompArx}. 

In highly isolated microscopic systems the environmental interaction rate may be much smaller, and the above separation need not hold. This is not a contradiction, since the present derivation concerns macroscopic or mesoscopic charges for which the environmental interaction represented by {\bf (RM)} is active. Thus the random-matrix mechanism and the Lorentz dynamics are compatible: the environmental interaction supplies rapid localization in the particle sector, while the ordinary charge-field Hamiltonian supplies the smooth classical force law between localization events. The field remains quantum throughout; classical electromagnetic behavior appears only in the localized quotient sector probed by the stabilized macroscopic charge.

\subsection{CMB radiation as a source of {\bf (RM)} dynamics}

The preceding subsection showed that the proposed classical field interpretation is compatible with the {\bf (RM)} stabilization mechanism in ordinary environments. We now consider a stronger question: whether, in the absence of air molecules or laboratory radiation, the cosmic microwave background itself can supply the environmental interactions with macroscopic bodies represented by {\bf (RM)}. If so, CMB radiation provides a universal source of localization for macroscopic bodies in otherwise empty space. 


At the temperature of the cosmic microwave background,
\[
T\simeq 2.7\,{\rm K},
\]
the photon density is much smaller than at room temperature, but it is not zero. For a body of radius
\[
R=10^{-6}\,{\rm m},
\]
the CMB photon incidence rate is of order
\[
\Gamma_{\gamma,{\rm in}}\sim 10^5\,{\rm s}^{-1},
\]
corresponding to
\[
\tau_{\rm env}\sim 10^{-5}\,{\rm s}.
\]
For the strong-field estimate
\[
\frac{Q}{M}=1\,{\rm C/kg},
\qquad
E=10^6\,{\rm V/m},
\]
the electromagnetic acceleration is
\[
a_{\rm EM}
=
\frac{Q}{M}E
=
10^6\,{\rm m/s^2}.
\]
The displacement due to this acceleration during one environmental time interval is
\[
\frac12 a_{\rm EM}\tau_{\rm env}^2
\sim
\frac12(10^6)(10^{-5})^2
=
5\times 10^{-5}\,{\rm m}.
\]
Thus, for a micron-sized body under these strong-field parameters, CMB scattering alone is not sufficient for localization at the scale \(10^{-7}\,{\rm m}\). It is, however, compatible with localization at the coarser scale
\[
\sigma\sim 10^{-5}\text{--}10^{-4}\,{\rm m}.
\]

For larger bodies the photon incidence rate grows as \(R^2\), and the corresponding environmental time scale decreases rapidly. For example, for a millimeter-scale body,
\[
R=10^{-3}\,{\rm m},
\]
the CMB photon incidence rate is obtained by scaling the micron estimate by the geometric cross section:
\[
\Gamma_{\gamma,{\rm in}}(R)
\sim
10^5
\left(
\frac{10^{-3}}{10^{-6}}
\right)^2
{\rm s}^{-1}.
\]
Thus
\[
\Gamma_{\gamma,{\rm in}}
\sim
10^{11}\,{\rm s}^{-1},
\qquad
\tau_{\rm env}
=
\Gamma_{\gamma,{\rm in}}^{-1}
\sim
10^{-11}\,{\rm s}.
\]
The electromagnetic displacement between environmental interactions is then only
\[
\frac12 a_{\rm EM}\tau_{\rm env}^2
\sim
\frac12(10^6)(10^{-11})^2
=
5\times 10^{-17}\,{\rm m}.
\]
Thus, for millimeter-scale macroscopic bodies, CMB photon scattering provides an environmental time scale much shorter than the Lorentz-motion time scale, even under the strong-field parameters used above.

The estimate above gives the physical time scale between environmental interactions supplied by CMB photon scattering. In the \({\bf (RM)}\) model, these interactions generate the random steps in projective state space. The return estimate is then the same as in the particle case. The relevant hitting behavior is governed by the Sparre Andersen theorem and is therefore insensitive to the detailed distribution of the individual random steps, provided the steps are independent and symmetrically distributed.

For the four-standard-deviation return estimate used in the particle case in \cite{KryukovPHLA}, the survival probability of a one-dimensional symmetric random walk satisfies
\[
P_{\rm no\,return}(N)
\sim
\frac{1}{\sqrt{\pi N}}.
\]
Thus, to obtain return probability
\[
P_{\rm return}(N)
\simeq
1-\varepsilon,
\]
one needs
\[
N_{\rm RM}
\sim
\frac{1}{\pi\varepsilon^2}.
\]
For the one-sided four-standard-deviation confidence level,
\[
1-\varepsilon\simeq 0.999968,
\qquad
\varepsilon\simeq 3.2\times 10^{-5},
\]
this gives
\[
N_{\rm RM}
\sim
\frac{1}{\pi(3.2\times 10^{-5})^2}
\sim
3\times 10^8 .
\]
The physical return time is therefore
\[
\tau_{\rm return}
\sim
N_{\rm RM}\tau_{\rm env}.
\]
For \(R=10^{-3}\,{\rm m}\), this gives
\[
\tau_{\rm return}
\sim
(3\times 10^8)(10^{-11}\,{\rm s})
=
3\times 10^{-3}\,{\rm s}.
\]
Thus, for millimeter-scale macroscopic bodies, CMB-induced \({\bf (RM)}\) stabilization gives a four-standard-deviation return time of order milliseconds.

This estimate separates two roles that electromagnetic radiation may play. Measuring devices, air molecules, thermal radiation, and other environmental degrees of freedom may all supply stochastic interactions represented by {\bf (RM)}. In such environments, even small or microscopic systems may be rapidly localized, as in a bubble chamber. The special role of the cosmic microwave background is different: it provides a universal background source of {\bf (RM)}-type interactions for macroscopic bodies even in otherwise empty space. Additional electromagnetic fields, whether laboratory fields or astrophysical fields on top of the CMB, then play a different role: they enter the Hamiltonian dynamics and determine the particular classical motion of the localized body. 

\section{Comparison with existing approaches}
\label{sec:comparison}

There are several standard approaches to the emergence of classical behavior in quantum field theory. The present construction is related to these approaches, but differs from them in its central object and in the role assigned to macroscopic probes.

A first class of approaches is based on decoherence and environment-induced superselection \cite{ZurekRMP2003,Schlosshauer2007,Joos2003}. In this picture, interaction with unobserved environmental degrees of freedom suppresses interference between macroscopically distinct alternatives and selects robust pointer variables. In quantum field theory and cosmology, this idea is often applied to field amplitudes or long-wavelength modes interacting with shorter-wavelength modes or other environmental degrees of freedom \cite{PolarskiStarobinsky1996,KieferPolarski2009}. Decoherence explains why interference between different field configurations becomes unobservable in the reduced density matrix. However, decoherence by itself does not identify a classical field with a localized quotient-level structure accessible to an individual macroscopic probe. In the present framework, decoherence-type suppression of interference may be compatible with the mechanism, but the classical field is represented instead by localized equivalence classes of field states selected operationally by macroscopic sources and probes.

A second standard route uses coherent states or more general semiclassical states \cite{Glauber1963,KlauderSkagerstam1985}. Coherent states are especially natural for harmonic systems and for free or weakly interacting fields, because their expectation values can follow classical equations with small relative fluctuations. In this approach, a classical field is often identified with the expectation value of the field operator in a coherent or semiclassical state. The present construction does not rely on such an identification. Gaussian field functionals may be used as convenient representatives for computing the local Fubini--Study geometry, but the physical object is the localized equivalence class of field states, not the coherent or Gaussian representative itself.

A closely related approach is based on Ehrenfest or mean-field equations. One studies expectation values of field operators and asks under what conditions these expectation values satisfy approximately classical field equations. Such methods are useful and often lead to correct semiclassical equations when fluctuations remain controlled. Nevertheless, expectation values of arbitrary quantum states do not by themselves define an observed classical field. The present approach differs in requiring localization in the observables accessible to macroscopic probes and in deriving the classical equations as the tangential component of the Schr\"odinger dynamics on the corresponding quotient manifold.

Another family of methods uses phase-space descriptions, Wigner functions, Wigner functionals, and coarse-grained distributions \cite{MrowczynskiMueller1994,Halliwell1987,HabibLaflamme1990}. These methods are well suited for studying semiclassical limits, quantum corrections to classical Liouville evolution, and classical statistical field approximations. They describe the transition from quantum dynamics to effective classical phase-space dynamics, often after coarse graining or in regimes where quantum interference terms are negligible. The present approach is not primarily a phase-space statistical construction. It is geometric: the classical field is represented by localized equivalence classes in projective Hilbert space, and the classical equations are obtained from the tangent component of the Schr\"odinger flow on the quotient manifold.

Finally, open-system and influence-functional methods derive effective equations for mean fields, correlation functions, or stochastic classical variables by integrating out environmental degrees of freedom \cite{CalzettaHu2008}. These methods naturally produce dissipation, noise, fluctuation--dissipation relations, and stochastic effective field equations. They are especially powerful in nonequilibrium quantum field theory and semiclassical gravity. In contrast, the present framework keeps the underlying dynamics unitary and uses the random-matrix mechanism {\bf (RM)} to stabilize macroscopic sources and probes near localized classical sectors. The role of {\bf (RM)} is not to replace field dynamics by an effective stochastic field equation, but to ensure that macroscopic particles remain localized and therefore probe only the quotient sector of the quantum field.

Thus the present construction is complementary to standard approaches. Decoherence explains suppression of interference, coherent-state and mean-field methods explain classical-looking expectation values, Wigner-functional methods describe semiclassical phase-space dynamics, and open-system methods derive effective noisy or dissipative equations. The new element here is the identification of the observed classical field with the quotient-level field data accessible to {\bf (RM)}-stabilized macroscopic probes. The quantum field itself remains quantum; classical field theory emerges as the tangential dynamics of localized equivalence classes selected by macroscopic interaction.

\section{Conclusion}

We have extended the geometric particle framework of \cite{KryukovPHLA,KryukovPhysicsA,KryukovArx,KryukovCompArx} to quantum fields. The central point is that the transition from a quantum field to an observed classical field does not require a change in the field itself, but only the macroscopic localization of the particles with which the field interacts. A classical field is represented by localized equivalence classes of field states, specified by the field data accessible to macroscopic sources and probes. The field remains quantum, while macroscopic particles stabilized near the classical particle manifold interact only with the quotient sector of the field state space visible through their finite spatial resolution. The resulting quotient-level field configuration is what is observed as a classical field.

The simulations illustrate this mechanism in finite-dimensional models. For the scalar field, a localized macroscopic probe selects field coordinates whose evolution agrees with the corresponding sourced Klein--Gordon equation. For the electromagnetic field, the quotient field coordinates satisfy Maxwell's equations. A localized test charge then responds to these same coordinates through the Lorentz force. Thus the field coordinates obtained from the quantum-field Schr\"odinger evolution are not merely formal solutions of classical field equations; they are the field data that determine the motion of localized macroscopic charges.

The geometric formulation explains why this happens. The localized field-state classes form a quotient manifold in projective Hilbert space. Classical field variables label these classes, and the classical field equations arise from the tangent component of the quantum-field Schr\"odinger dynamics on this manifold. Gaussian field functionals are useful representatives for computing the induced Fubini--Study geometry, but the construction is not a coherent-state or special-state ansatz. The physically relevant object is the localized equivalence class, not the particular representative used to parametrize it.

The role of the random-matrix mechanism {\bf (RM)} is complementary. It does not replace the Hamiltonian dynamics responsible for classical field equations or the Lorentz force. Instead, environmental interactions represented by {\bf (RM)} stabilize macroscopic particles near the localized particle manifold, making them effective classical sources and probes of the field. Measuring devices, natural environments, and radiation fields may all contribute to this stabilization. The cosmic microwave background provides a universal limiting source of such interactions for macroscopic bodies in otherwise empty space, while stronger local environments may produce much faster localization, including in measurement situations involving microscopic particles.

The resulting picture is that classical field theory emerges from quantum-field dynamics only relative to stabilized macroscopic probes. The quantum field itself is not reduced to a classical object; rather, the classical field is the quotient-level structure of the quantum field that is operationally accessible to macroscopic localized systems. In this sense, the tangent dynamics of localized equivalence classes provides a unified geometric route from unitary quantum dynamics to classical particle and field dynamics.

\section{Declaration of interest statement}

The author declares no competing interests.


\end{document}